# The Importance of Cognitive Domains and the Returns to Schooling in South Africa: Evidence from Two Labor Surveys[★]

Plamen Nikolov[★,a,b,c,d]    Nusrat Jimi[a]

**Abstract**. Numerous studies have considered the important role of cognition in estimating the returns to schooling. How cognitive abilities affect schooling may have important policy implications, especially in developing countries during periods of increasing educational attainment. Using two longitudinal labor surveys that collect direct proxy measures of cognitive skills, we study the importance of specific cognitive domains for the returns to schooling in two samples. We instrument for schooling levels and we find that each additional year of schooling leads to an increase in earnings by approximately 18-20 percent. The estimated effect sizes—based on the two-stage least squares estimates—are above the corresponding ordinary least squares estimates. Furthermore, we estimate and demonstrate the importance of specific cognitive domains in the classical Mincer equation. We find that executive functioning skills (i.e., memory and orientation) are important drivers of earnings in the rural sample, whereas higher-order cognitive skills (i.e., numeracy) are more important for determining earnings in the urban sample. Although numeracy is tested in both samples, it is only a statistically significant predictor of earnings in the urban sample. (JEL I21, F63, F66, N37)

___________________________

[★]We are grateful to the research staff at the Harvard Center for Population and Development Studies who made the HAALSI data available to us and provided numerous insights based on their field experience implementing the survey. We also thank Jerray Chang for his research assistance and invaluable input in the early stage of our study. Plamen Nikolov gratefully acknowledges research support by the Economics Department at the State University of New York (Binghamton) and the Research Foundation for SUNY at Binghamton. We thank Matthew Bonci, Declan Levine, David Titus, and Steve Yeh for outstanding research support with this project. We thank Susan Wolcott, Christopher Hanes, Nusrat Jimi, Eric Edmonds, Alan Adelman, Xu Wang, Subal Kumbakhar, Zoe McLaren for constructive feedback and helpful comments. All remaining errors are our own. All remaining errors are our own.

[a] Department of Economics, State University of New York, Binghamton
[b] Harvard University Institute for Quantitative Social Science
[c] IZA Institute of Labor Economics
[d] Global Labor Organization
[e] University of Pennsylvania

## I.  Introduction

Education is widely accepted as a leading instrument for promoting better economic outcomes at an individual level (UNESCO 2005). Following the innovative analyses by Mincer (1958, 1974), the literature examines the economic returns to differing levels of schooling at an individual level.[1] Education can positively influence non-monetary outcomes, as it can reduce crime (Lochner and Moretti 2004), improve health (Currie and Moretti 2003), and improve civic engagement (Milligan et al. 2004). A large body of empirical literature investigates the monetary returns to education in developed and developing countries.[2] In general, estimated returns to education are larger in developing countries than in industrialized countries. Previous studies that examine the returns to education in developing countries rely on observational study designs. For sub-Saharan Africa, where rates of poverty are among the highest in the world (World Bank 2016), education is particularly important. The issue of causal estimates of the returns to schooling is central for careful policy targeting. However, previous empirical estimates that used data from sub-Saharan Africa rely on observational study designs, which are ill-equipped to detect true causal effects. Precisely measuring the returns to schooling in the context of sub-Saharan countries is a challenge plagued by econometric issues, data availability constraints (Card 1999, 2001) and survey design issues (Serneels et al. 2017).[3,4,5] For example, the socioeconomic characteristics of households and communities are important determinants of both schooling and labor market outcomes in developing countries. The omission of individual ability measures in empirical estimations can lead to large upward bias (ability bias) of observational study estimates (Lang 1993). In practice, the bias present in estimates based on observational study designs could be quite large.

---

[1] See Psacharopoulos (1994), Card (1999), Harmon et al. (2003), Psacharopoulos and Patrinos (2004), and Heckman, Lochner, and Todd. (2006). Lin, Lutter and Ruhm (2018) examine how cognitive performance relates to labour market outcomes in the U.S; Walsh (2013) examines the effect of ability proxies on earnings in Ireland.

[2] For surveys of empirical estimates in developed countries, see Card (1999) and Harmon et al. (2003). For empirical estimates in developing countries, see Psacharopoulos (1994), Patrinos and Psacharopoulos (2010) and Nikolov and Jimi (2018). Chakravarty et al. (2019) review estimates in the context of vocational schooling in developing countries.

[3] Card (2001) and Duflo (2001) demonstrate that returns to schooling are typically higher in developing countries than in developed countries. However, estimates on the returns to schooling in developing countries is scarce. Prior to Card (2001) and Duflo (2001), estimates on the returns to schooling for low and middle-income countries were largely based on Patrinos and Psacharopoulos (2010).

[4] Using a field experiment in Tanzania, Serneels et al. (2017) examine whether variation in survey design introduces measurement error when estimating the returns to education. The study shows that, although the estimated returns to education vary by questionnaire design, they do not vary by whether information on wages is self-reported or collected by a proxy respondent.

[5] All methods have some shortcomings, as discussed by Card (2001) and Heckman et al. (2006). Furthermore, estimation of the returns to schooling in developing countries is even more challenging given the scarcity of rich datasets.

In this paper, we examine the returns to cognitive skills and the returns to schooling using data from South Africa.[6] We use two surveys, which collected extensive household information including various dimensions of cognition. The surveys were conducted in South Africa between 2002 and 2014 in two distinct regions of South Africa: urban and rural. They provide data on how various dimensions of cognition influence schooling and labor market outcomes in rural and urban settings. In particular, the use of two distinct surveys from two contrasting areas of South Africa enables an examination of the implications of heterogeneous local labor markets on the importance of specific cognitive domains. The first survey that we use is Health and Aging in Africa: A Longitudinal Survey of an INDEPTH Community in South Africa (HAALSI). HAALSI examines a rural population aged 40 and older. The second survey, the Cape Area Panel Study (CAPS), follows a metropolitan area population aged between 14 and 22. Both surveys collect cognitive evaluations via direct measurements of specific cognitive domains. In HAALSI (2014), four cognitive domains were tested: memory, numeracy, orientation, and attention. In CAPS (2002–2009), a cognitive evaluation was administered in Wave 1 (2002) and evaluated the domains of literacy and numeracy. Using data from the five waves of the CAPS panel data and the baseline wave of the 2014 HAALSI, we examine the influence of each cognitive domain on reported earnings. Furthermore, we use principal component analysis (PCA) to compile information from specific cognitive domains into an aggregate index and examine the impact of both specific cognitive measures and the aggregate cognitive index on the estimated returns to schooling.

Theoretical research by Ben-Porath (1967) and Becker (1975, 101, n. 89), and empirical studies in sub-Saharan countries (Boissiere et al. 1985; Field et al. 2009), indicates that cognitive skills influence schooling attainment. The failure to account for measures of cognitive ability in observational study designs that estimate the Mincer equation tends to bias the return to schooling estimates. Recent empirical comparisons have shown a gap between ordinary least squares (OLS) and instrumental variables (IV) estimations of the returns to schooling (Card 2001; Trostel et al. 2002). The magnitude of ability bias in the returns to schooling can be an important factor that contributes to this discrepancy. Without causal estimates of the returns to schooling, gauging the ability bias in observational study designs can shed some light on the true lack of accuracy of observational study estimates on the returns to schooling in low-income countries.

---

[6] To some extent, this paper can inform debates about ability bias in observational study designs. It is important to add the caveat that cognitive skills are an important proxy for ability, but they are not a direct measure of ability, which is a multidimensional concept.

We report three major findings. First, we find that the OLS estimate for the return to each additional year of schooling acquired is 14 percent (based on Wave 5 of the CAPS). Based on HAALSI, the OLS estimate for each additional year of schooling acquired is approximately 10 percent. These estimates are similar to estimates based on survey data from sub-Saharan Africa from Psacharopoulos and Patrinos (2004). Second, when we instrument for individual schooling levels using schooling fees, we find evidence consistent across all two-stage least squares (2SLS) specifications of higher effect size. Using the schooling fees as an instrument, we find that each additional year of schooling leads to an increase in earnings by approximately 18-20 percent (significant at the 5-percent level). When we use the quarter of birth as an instrument, the estimated return to each additional year of schooling increases up to 28 percent but these estimates are imprecisely estimated.[7] Our IV estimates are generally higher than the OLS estimates, a difference likely due to the 2SLS estimation being based on individuals with high marginal returns to schooling. In the third estimation, we detect evidence that suggests that specific cognitive domains play a far more important role in determining a person's earnings than general cognitive skills.[8] Virtually no prior research examines the importance of cognitive skills for earnings in sub-Saharan countries. This is largely due to a lack of longitudinal data on specific cognitive skills and earnings. Based on the two samples, the overarching pattern that we observe is that, in the rural sample, executive functions (e.g., memory or orientation skills) are more important than earnings. Conversely, in the urban sample, higher-order cognitive skills are more important. Although numeracy skills are evaluated in both samples, it is a statistically significant predictor of earnings only in the urban CAPS sample.

---

[7] We examine for any potential seasonality patterns of births that might threaten the validity of the 2SLS approach based on the quarter-of-birth instrument. When we examine for seasonality of births, specifically in months related to the seasonality of food supply, we detect no evidence of seasonality of births in the sample.

[8] We implement robustness tests to examine whether proxy measures for ability are influenced by a schooling gradient, and we detect no evidence that they are influenced by schooling differences.

This paper makes four important contributions to the labor economics literature on the returns to schooling in developing countries. First, we contribute to understanding the effects of cognitive skills on the returns to education. We estimate and demonstrate the importance of specific cognitive domain proxies in the classical Mincer equation. Our results suggest that cognitive ability scores explain a sizable, positive effect on earnings and slightly diminish the effect of schooling on earnings. Related to this issue, we show that the earnings gap between blacks/coloured and whites in South Africa is considerably reduced when we account for the ability measures. Neal and Johnson (1996) find a similar result, in the context of the U.S. racial wage gap. In contrast, cognition measures do not influence the gender earnings gap. These findings imply that differences in premarket skill do account for a significant portion of the earnings gap across population groups but do not seem to play a large role in the gender earnings gap in South Africa. Second, we contribute to the existing empirical literature by estimating the returns to certain dimensions of cognition that proxy for innate ability. Furthermore, cognitive skills measures implicitly provide information about the fundamental role of schooling quality and non-school factors in the production of human capital, which have been overlooked in the previous standard estimations of returns to schooling (Hanushek and Woessmann 2008). Our results suggest that certain domains of cognition may be more important for earnings than other domains are. When we split the ability proxy into specific cognitive domain components, we find that only the numeracy domain in the CAPS sample and the memory and orientation domains in the HAALSI sample are statistically significant determinants of earnings. Third, although we rely on a limited sample size and an arguably imperfect instrument, we provide suggestive evidence of the causal estimates of the returns to schooling in South Africa by exploiting an IV estimation approach. We also provide suggestive evidence on the magnitude of the ability bias in the context of a developing country. Although much has been written about ability bias (Lang 1993), and numerous empirical papers have estimated arguably causal returns to schooling in high-income countries (Card 2001), little is known about the true casual returns in the context of low-income countries, in sub-Saharan Africa, in particular. Although the measures of cognitive skills that we use are an important proxy for ability, they are not a direct measure of ability. Nevertheless, we examine how the inclusion of proxies for ability enhances understanding of the effects of cognitive skills on the returns to education. Finally, we provide additional evidence on the returns to schooling based on two survey sources, drawn from two demographic groups. HAALSI (conducted in 2014) surveyed an aging generation of rural South Africans, aged 40 and older. In contrast, the CAPS, collected between 2002 and 2009, gathered information on young adults, aged between 14 and 22, in a metropolitan area. By using two distinct samples—one based in an urban setting and another based in a rural setting—with rich data on various cognitive domains, we gain a better understanding of the relationship between the returns to schooling and specific cognitive domains in two distinct settings in the context of a developing country.

The remainder of the paper is structured as follows: Section II provides background information about South Africa's education system, major cognitive domains, and previous research about the role that cognitive domains play in determining a person's earnings. Section III describes the survey data, sampling strategies, and overall sample characteristics. Section IV presents the empirical strategy. Section V reports the results. Section VI describes various robustness checks. Section VII offers a discussion and concluding remarks.

## II. Background

### A. The Education System in South Africa

*Educational Levels and Compulsory Schooling Requirements*. In South Africa, school spans 13 years or grades: from grade 0, known as grade R, through to grade 12 or the year of matriculation (the "matric"). The education system comprises three tiers: general education and training, further education and training, and higher education and training. Table 1 presents information on how years of schooling relate to grade levels in the educational system. General education and training run from grade 0 to grade 9. The first six years are spent in primary school, where students acquire literacy and numeracy skills. After that point, students enter the further education and training phase (grades 10 through 12). Under the South African Schools Act of 1996 (Davel 2000; Maithufi 1997; Sayed and Jansen 2001), education is compulsory for all South African people from the age of seven

(grade 1) to 15, or the completion of grade 9.[9,10,11]

[Table 1 about here]

*The Apartheid System and the Legacy of Racial Segregation.* The education system was also marked by the legacy of the apartheid, which began when the National Party won control of South Africa's Parliament in 1948 and racial segregation in education was a major pillar of the political regime. The 1953 Bantu Education Act brought African education under the control of the government and extended apartheid to black schools. In 1963 and 1964, a separate education system was established for the education of "coloureds" and Indian people.[12] An education act for white people was passed in 1967; glaring inequalities existed between the race-based schooling systems in South Africa. These inequalities applied to teacher qualifications, teacher–pupil ratios, funding per student, facilities, books.[13,14] The race-based education system was only fully dismantled in 1994, although, toward the end of this period, some modest concessions were made that allowed white schools to enroll some black children. Issues of race and racial discrimination continue to plague the education system, affecting access to school and the demand for schooling among South Africans (Mwabu and Schultz 1996; Salisbury 2016; Schultz 2004).

*Taxonomy of Cognitive Domains.* Since a major focus of our analysis will be on cognitive skills and domains, as they relate to educational attainment and a person's earnings, we proceed with a brief overview of the cognitive domains that are critical to economic outcomes. Numerous fields of social science and science have studied the brain and its functions. Each of the brain's roles—movement, sensory input, and interpretation—are important for enabling a person to function in their daily life.[15] Although there is general understanding and agreement about the various cognitive functions, there is no single way to categorize the various components of cognition. The four major domains are attention, inhibitory control, memory, and higher-order cognitive functions.

---

[9] South Africa relies on the matric pass rate as a significant marker of the progress of its schools.
[10] Education is compulsory only during the general education and training phase.
[11] The National Senior Certificate is the benchmark secondary education qualification in South Africa, which is awarded by Umalusi (Council for Quality Assurance in General and Further Education and Training) upon successful completion of school and the grade 12 examination. Students who are more vocationally or technically oriented may choose to work on N1, N2, and N3 Certificates in a particular vocation or craft instead.
[12] "Coloured" is a term used to describe a person of mixed European (white) and African (black) or Asian ancestry and is officially defined by the National Statistical Bureau of South Africa (Statistics SA) (Statistics SA 2017).
[13] Eight education departments followed different curricula and offered different standards of learning quality for primary and secondary schooling. As a result, white schools fared the best; Indian and coloured schools fared better than those for black Africans did (Morrow 1990).
[14] The poor quality of education during apartheid was particularly disadvantageous for black South Africans. Black people were placed in poorly financed public schools (Crouch 1996; Kriege et al. 1994; Marais 1995) that were of poorer quality (Case and Deaton 1999; Case and Yogo 1999; Card and Krueger 1992). Crouch and Lombard (2000) report notable discrepancies between racial groups on exam scores, particularly in numeracy.

The first major domain is attention. Attention is the ability to focus on information by engaging in a filtering process of the incoming stimuli. Processing stimuli can happen either voluntarily or involuntarily. For example, attention alerts a person to loud music (involuntarily) or enables them to comprehend a sentence within a book (voluntarily). The second major domain is inhibitory control—that is, a person's ability to control incoming temptations and minimize interference from stimuli that are deemed irrelevant. This inhibitory control is effective for blocking a series of specific stimuli such as distractions, strong and sudden urges, and pre-potent responses. The third major domain is memory. Memory is the ability to retrieve, recognize, and use information that has been previously learned. There are various types of memories: sensory, short-term, and long-term. Sensory memory is the ability to retain impressions of sensory information once the original stimuli have ended. Short-term memory is a person's ability to assess and manipulate new information. Long-term memory is the ability to store information over a long period. The fourth major cognitive domain is higher-order cognitive functions. These functions involve one or more of the basic cognitive functions discussed above and, because they rest on simpler processes, this domain is inherently more complex. There are three major higher-order cognitive functions: cognitive flexibility, intelligence, and planning. Intelligence is of particular interest to this study, as many cognitive tests attempt to measure one or more facets of intelligence. Generally, psychologists have separated intelligence into crystallized and fluid intelligence (Cattell 1963). Crystallized intelligence involves a person's ability to use learned skills or languages, whereas fluid intelligence refers to a person's ability to solve real-world problems.

---

[15] A related strand of studies examines the way cognitive skills or domains map to school performance, educational attainment and earnings. In the education literature, Bloom (1956) and Anderson et al. (2001) highlight a taxonomy of hierarchical levels of thinking and cognitive skills for a person's educational success. In the psychology literature, studies point to various facets of cognition as predictors of quality of learning (Almeida et al. 2008; Spinath et al. 2006; Sternberg 2012; Sternberg et al. 2001; Strenze 2007) and are important factors that influence a person's academic achievement (Deary et al. 2007; Howes et al. 1999; Kurdek and Sinclair 2001; Lemos et al. 2011; Primi et al. 2010; Riding and Pugh 1977). A recent report from the University of Chicago argues that the noncognitive skills that are most strongly associated with academic performance are academic behaviors (e.g., attending class and participation), academic perseverance (e.g., grit and self-discipline), academic mind-set (e.g., feeling a sense of belonging in an academic community and believing that ability and competence can grow with effort), learning strategies (e.g., meta-cognitive strategies and goal setting), and social skills (e.g., interpersonal skills and cooperation) (Farrington et al. 2012; Kautz et al. 2014).

## B. Cognitive Skills and Their Influence on Labor Market Outcomes

Numerous studies have examined the positive effects of cognitive performance on labor market outcomes in high-income countries (Hanushek and Woessmann 2008; Heckman 1995; Murnane et al. 1995, 2001; Neal and Johnson 1996).[16,17] Two related strands of literature examine the relationship between cognitive performance and economic performance. In one strand, studies examine the importance of cognitive skills for specific sectors and occupations. For example, Ree et al. (1994) and Guion (1983) document the importance of cognitive skills in industries and occupations that rely on relatively technology-intensive inputs. A more crowded strand of literature, predominantly based on the experiences of middle-income and high-income countries, examines returns to cognitive skills (Bowles et al. 2001).[18] Building on Bowles et al. (2001), subsequent studies, such as Heckman, Stixrud, and Urzua (2006), Lindqvist and Vestman (2011), Lin et al. (2018), and Barone and Van de Werfhorst (2011), examine the effects of cognitive skills on labor market outcomes.[19]

Most of the empirical evidence on the importance of general IQ or general cognitive skills on economic outcomes comes from data from high-income countries examines. This literature suggests a substantial premium for cognitive skill in the U.S. labor market (Bowles et al. 2001; Howell and Wolff 1991; Murnane et al. 1995, Denny and Doyle 2010; Heineck and Anger 2010; Vignoles et al. 2011; Barrett 2012; Ramos et al. 2013). In the context of a low-income country, only LaFave and Thomas (2017), who use data from Indonesia, examine the effects of various cognitive domains (i.e., Raven's test, fluid intelligence, and memory) on earnings.

---

[16] Follow-up debates focused on genetic versus environmental factors and questioned whether cognitive ability can be characterized by a single measure. Heckman (1995) points out that no agreement exists regarding the use of a single latent ability factor to explain the data or the existence of only one type of cognition. Hanushek and Woessmann (2008) review the role of cognitive skills in influencing economic well-being, with a special emphasis on the role of the quality and quantity of schools.

[17] Glewwe et al. (2017) examine the effect of cognitive skills on labor market outcomes in rural China and find no strong evidence that earlier skills predict wages in the early years of labor market participation.

[18] A well-known challenge that arises when estimating the returns to cognitive performance is that observed relationships need not reflect causal effects, as a constellation of other factors may have independent confounding effects on labor market outcomes. Heckman et al. (2006), and subsequent work, show that non-cognitive skills have large returns in the labor market. In the context of high-income countries, previous studies document the influence of cognitive skills on various labor market outcomes; that is, they have previously focused on weekly earnings (e.g., Griliches and Mason 1972), annual income (e.g., Salkever 1995), and hourly wages (e.g., Neal and Johnson 1996).

[19] A growing body of empirical studies documents the importance of non-cognitive skills for labor market outcomes (Borghans et al. 2014; Heckman and Kautz 2012; Heckman, Stixrud, and Urzua 2006; Kuhn and Weinberger 2005; Lindqvist and Vestman 2011).

## III. Survey Data and Measures of Cognitive Domains

To examine the magnitude of the ability bias on the returns to schooling using micro-level estimates from South Africa, we use two primary data sources: the 2014 to 2015 HAALSI and the 2002 to 2006 CAPS. The CAPS followed young adults in a metropolitan area and HAALSI tracked adults aged 40 and above in the rural Mpumalanga province of South Africa.

### A. Survey Data

*The Health and Aging in Africa Longitudinal Survey.* Our first data source is HAALSI, a survey funded and conducted by the Harvard Center for Population and Development Studies and a sister survey of the U.S.-based Health and Retirement Survey (HRS). It studies the aging of men and women in the rural province of Mpumalanga in South Africa. The sample includes approximately 5,000 men and women aged 40 and older who lived in the Agincourt Health and Socio-Demographic Surveillance System site during the 12 months prior to the 2013 census.

The primary objective of the survey was to identify determinants of ill-health and disability and their effects on subjective well-being, household composition, household income and expenditure, labor force participation, and physical and cognitive functioning. Baseline survey interviews were conducted between November 2014 and November 2015. The random sample of respondents had a response rate of 87 percent (Harling et al. 2018).[20]

In addition to surveying the basic socioeconomic information[21,22] of participants, HAALSI includes a cognitive test based on the cognitive module of the U.S.-based HRS. HAALSI covers three major domains: numeracy, memory, orientation, and numeracy. The numeracy portion contains a counting test and a number pattern recognition test. The memory portion comprises a two-part word memorization test and a self-rated memory test. A self-rated attention assessment is also included.[23]

---

[20] Those who refused to participate were slightly more likely to be women and were more likely to be native South Africans (nationality). The population group was not captured in HAALSI, as black Africans almost exclusively populated the survey's geographic coverage.
[21] HAALSI collected individual-level information about current employment, wage rate in current employment, or employment in the past 30 days, hours worked, and type (sector) of job.
[22] Serneels et al. (2017) note the importance of questionnaire design for accurately reporting earnings in the context of developing countries. They note that a long questionnaire design is usually preferable. Both the CAPS and HAALSI employ detailed questionnaires for their labor survey modules.

*The Cape Area Panel Study.* Our second data source is the CAPS. The CAPS tracked young adults and older adults from 2002 to 2009. The primary focus of the CAPS was on young adults aged between 14 and 22 in the metropolitan Cape Town area. The panel sample of young adults comprises 4,752 individual surveys and 46 percent of the survey sample was male.

The first of four waves of this survey comprised 5,291 randomly selected young people in late 2002. With the anticipation of an 80 percent response rate, the completed interviews aligned with the desired sample size of roughly 4,800.[24,25] In 2003, one-third of the original sample was re-interviewed. The remaining two-thirds of the sample were revisited in 2004. The entire sample was re-interviewed in 2005 and again in 2006. Finally, interviews over the phone in 2009 provided data for a fifth wave. As the primary purpose of this study is to measure the returns to schooling, we primarily consider earnings data from Wave 5, where the greatest number of young adults in the sample are old enough to work.

The CAPS included a cognitive evaluation (the literacy and numeracy evaluation) along with schooling, earnings,[26] and socioeconomic data. The evaluation contained 45 questions that were evenly split between literacy and numeracy assessments. It is important to note that, although the evaluation was given in either English or Afrikaans, some individuals primarily spoke Xhosa. Of Xhosa native speakers, 99 percent chose to take the

---

[23] The tests and questions used in the cognitive evaluation are widely supported by psychological and epidemiological research (Harling et al. 2018).

[24] The sampling design of the CAPS comprised two stages. The first was the selection of clusters using the 1996 Population Census. The second stage was the selection of households within those clusters. The 1996 Population Census categorized geographic locations with enumeration areas (EAs). To categorize EAs for selecting different population groups, the survey used the 10 percent micro-data from the 1996 Census to identify the number of households belonging to the different population groups by EA. In the second stage, households were selected using aerial photographs of each EA and assigned a number. Like many surveys, low numbers of responses to earnings information limited the number of observations that were useful for analysis. CAPS includes sampling weights to allow adjusting for nonresponse bias for each wave; non-responsiveness can change the representativeness of the sample population.

[25] The actual household response rate was 74 percent. The individual response rate for the CAPS was 89.6 percent for young adults conditional on the household having been interviewed. Differences across population groups were much smaller than they were for overall household participation. Response rates for young adults were 93 percent for African young adults, 88 percent for coloured young adults, and 86 percent for white young adults. At the individual level, older respondents tended to have lower response rates although a comparison between the characteristics (mean) of respondents and non-respondents was not reported.

[26] The CAPS Labor Module was conducted at an individual level; it collected information on the respondent's labor force activity in Waves 1, 2, 3 and 4.

English version of the literacy and numeracy evaluation, which evaluated their English language skills as well as cognition.

**B.    Sample Summary Statistics**

Table 2 provides summary statistics for HAALSI, the CAPS (Wave 5), and data on the same socioeconomic and demographic variables except for the general population from the Quarterly Labour Force Survey (QLFS).[27][28] The QLFS is representative of the South African population and is based on South Africa's General Census. Table 2 reports summary statistics from the QLFS to compare the characteristics of the CAPS and HAALSI samples and the general population of South Africa. The two surveyed areas differ substantially in ethnic makeup, the average age of participants, geographic setting, and urbanicity.[29] The rural, northeast sub-district of Agincourt (where HAALSI was administered) has higher poverty rates, lower educational attainment rates, higher unemployment rates, and higher labor migration patterns compared with the location in the CAPS. For the population surveyed by HAALSI, the average age is 54.51 years and the average years of schooling is 3.68. The racial makeup of Agincourt is almost 100 percent black African, although this is not shown in Table 2.

[Table 2 about here]

Table 2 also reports summary data based on the QLFS on two relevant nationally representative groups: summary statistics of individuals aged from 21 to 29 (reported in column 4) and summary statistics of individuals aged 40 and above (reported in column 8). These groups provide a comparison of the nationally representative survey samples and the two survey samples on which we base our analysis.[30]

---

[27] Appendix Table A1 reports summary statistics for the CAPS as a panel by individual wave.
[28] The QLFS sample is based on information collected during the 2001 Population Census implemented by Statistics SA. The QLFS sample was designed to be representative at a provincial level and within provinces at a metro and non-metro level.
[29] In contrast to the HAALSI population, the CAPS (Wave 5) population has an average age of 24.88 years and an average schooling attainment of 10.76 years.
[30] A comparison of the summary data reported in column (4) (based on the QLFS data) and column 1 shows that the CAPS sample tracks the pattern of the general population in terms of demographic, social, educational, and economic characteristics. One noticeable difference is that the black African population is slightly overrepresented in the CAPS sample compared to the relative percentage of that population group in the QLFS urban subsample.

## C. Cognitive Ability Measures in the Surveys

In addition to collecting household socioeconomic information, both the CAPS and HAALSI collected information on the various domains of cognition (summarized in Appendix Table A2).

In particular, the CAPS measured two domains: literacy and numeracy. The survey instrument included two modules (the literacy and numeracy evaluation modules of the CAPS Wave 1 questionnaire) to quantify these cognitive domains. The evaluation instrument comprised 45 questions relating to literacy and numeracy. For the literacy task, respondents were asked to complete a series of tasks. They were asked to match a word with a visual image that matched the written word. Another task was to choose a word that logically completed a sentence. The final literacy task was to read a passage and respond to various reading comprehension questions about the text. The numeracy score was based on a series of questions. For example, respondents were asked to arrange numbers from smallest to largest. They were also asked to complete logic-related exercises that relied on their ability to count, add, or subtract numbers. Based on the questions in these domains, two aggregate scores (one for literacy and one for numeracy) were created.[31]

HAALSI also captured cognitive measurements. The cognition module of HAALSI relates to four domains: memory, attention, orientation, and numeracy.[32,33] Additionally, the instrument included questions to capture self-rated memory and attention.[34] Memory was measured using a series of orientation questions. Interviewers audio-recorded answers and scored the response accuracy for each item immediately after each response was given. Respondents scored one for each correct answer, with a maximum of four. For the domain of attention, in addition to self-rated attention questions, respondents were given an auditory attention task that assessed selective and sustained attention to words that were repeatedly presented in an audio recording. For the orientation domain, respondents were asked to name the month, date, year, and president. The accuracy of the responses (correct, incorrect, not known, or refusal) was recorded for each question. Numeracy was measured by various tasks in which the respondent was asked to add, subtract, or use logical skills related to number manipulation. Similar to the procedure used for CAPS, we aggregated information from the survey questions that related to each cognitive domain into an index that captures a cognitive domain index based on a score that used PCA.

---

[31] Based on the aggregate scores, we created one aggregate score using PCA. We accomplish the reduction by estimating weighted linear combinations that contain the cognitive domain measures.

[32] Both the CAPS and HAALSI measured a domain associated with numeracy skills. In CAPS, numeracy skills were measured by asking respondents to complete simple algebraic tasks and logic-related exercises that relied on the person's ability to count, add, or subtract numbers. HAALSI captured numeracy skills by asking respondents to count backward and conduct simple computations involving basic mathematical operations such as addition, subtraction, division, multiplication, and percentages. Additionally, HAALSI implemented a common numeracy test called the serial seven task, in which the respondent was asked to subtract seven from 100 in the first step and then continue to subtract seven from the previous number in each subsequent step.

[33] CAPS has a distinct survey domain that measures literacy. HAALSI measures three distinct domains: memory, orientation, and attention. With the exception of numeracy skills, the two surveys measure distinct domains; however, we can gain important insights about the returns to specific cognitive domains from the two surveys.

[34] As in the study by Griliches (1977), the cognitive evaluations are not ideal because they are conducted after some schooling has been completed. This could suggest that the wage effect of ability in the returns to schooling estimation is a function of schooling. We address this possibility with a robustness check in our analysis. Although the cognitive ability evaluation may already reflect the earning or schooling trajectory of students by the time the CAPS Wave 1 is conducted, it is unknown how early a cognitive assessment might be necessary to eliminate major interference from education. Additionally, Laajaj and Macours (2017) find cognitive tests consistent and reliable in developing countries. In HAALSI, the cognitive evaluation was designed to measure cognitive performance to indicate symptoms of aging and was not intended for assessing ability as it relates to earnings. An evaluation designed for economic and educational research would be optimal, but no such survey exists in South Africa. Nonetheless, the cognitive assessment was based on an international standard set of questions established by the HRS, which supports the validity of the cognitive domain specification. Furthermore, this strengthens the feasibility of using a proxy for ability in the returns to schooling estimation.

## IV. Empirical Strategy

### A. Returns to Schooling: OLS Approach

The objective of this paper is to estimate the returns to schooling using the two labor surveys, with and without survey measures of cognitive development. Following a long tradition in labor economics based on Mincer (1974) and Heckman, Lochner, and Todd (2006), we estimate a standard Mincer equation using the following specification:

(1) $\quad \ln y_i = \beta_0 + \beta_{1i} S_i + \beta_{2i} exp_i + \beta_{3i} exp_i^2 + \beta_{4i} Gender_i + \beta_{5i} PopGroup_i + \varepsilon_i,$

where $y_i$ is earnings (monthly earnings reported in the previous month). For each individual $i$, $S_i$ captures a person's schooling attainment (measured in number of years) and $exp_i$ and $exp_i^2$ capture a person's experience and work experience squared.[35] We include a binary dummy variable for gender (1 if male). Finally, following the population group definitions for South Africa's Census Survey, we also include two binary indicators for population group: Black (1 if Black African) and Coloured (1 if Coloured)[36]; the reference population group is White. $\beta_1$ captures one's rate of return to schooling.

---

[35] Experience is determined by differencing age and schooling minus six, following Mincer (1974), Boissiere et al. (1985), and Lemieux (2006).

A key problem with measuring the returns to schooling through the earnings–schooling function is that schooling decisions are often not truly exogenous. An individual decision about how much schooling to obtain are likely to also correlate with one's unobserved characteristics. Other factors, such as community background characteristics or the person's innate ability, are likely to have an important influence in one's schooling decision. In the Mincer equations, the omission of variables that capture dimensions of one's ability will lead to an upward bias of the estimate of the return to schooling. These estimations are further susceptible to measurement error and measurement error bias. Individuals often misreport earnings data and precise schooling attainment can be difficult to track in developing countries. Card (2001) points out that OLS estimates of the returns to schooling that rely on observational study designs are largely downward biased, despite ability leading to upward bias of the rate of return to schooling coefficient.

We estimate specification (1) using the CAPS data in two ways: using Wave 5 only and using a panel analysis of Waves 1, 3, 4, 5. The panel analysis is based on the observations with continuous earnings data throughout Waves 1, 3, 4, and 5 and it enables us to estimate the Mincer equation using the panel data structure.

We similarly estimate specification (1) using HAALSI baseline data. The earnings data for HAALSI derive from two variables in the HAALSI questionnaire: current job earnings and last job earnings. All earnings data are adjusted for inflation based on the year the individual was last working. We adjust earnings data to constant 2002 ZARs for all earnings data. To accurately measure the experience variable for individuals who are retired or not working, we transform the age used in the analysis to the age when the individual was last working.[37]

In addition to specification (1), we also estimate the returns to schooling by including measures of cognitive ability:

---

[36] We use the population group definitions for South Africa's Census Survey. The major population groups are black African, coloured, and white. The reference population group in this specification is white.
[37] Without adjusting age, individuals who have been retired for many years would have inaccurately large experience values in relation to their wage, which is likely to create bias in the estimates in our study.

(2)     $\ln y_i = \beta_{0i} + \beta_{1i} S_i + \beta_{2i} exp_i + \beta_{3i} exp_i^2 + \beta_{4i} Gender_i + \beta_{5i} PopGroup_i + \beta_{6i} A_i + \varepsilon_i,$

where $A_i$ is one's cognitive ability index based on survey measures. We also estimate (2) using specific cognitive ability domains by including each specific cognitive ability domain ($A_{ij}$) as a separate measure:

(3)     $\ln y_i = \beta_{0i} + \beta_{1i} S_i + \beta_{2i} exp_i + \beta_{3i} exp_i^2 + \beta_{4i} Gender_i + \beta_{5i} PopGroup_i + \sum_{i,j=1}^{n} \beta_{6,j} A_{ij} + \varepsilon_i,$

where $A_{ij}$ is a matrix of $j$ cognitive domain components that are measured as an index for each individual. The final specification provides a measure of the significance and magnitude of each cognitive domain to explain earnings differentials.

However, it is necessary to exercise caution when interpreting the coefficients associated with the cognitive measures from the specifications above, as the returns to experience and education do not necessarily reflect returns to accumulated human capital. Because we are interested in labor market rewards to cognitive skills, our goal is to estimate the direct effect of variations in cognitive performance on earnings, not the indirect effects operating via the effects of cognitive scores on the level of schooling attained with the returns to schooling.[38]

### B.     Returns to Schooling: 2SLS Approach

To tackle the endogenous nature of the schooling variable, numerous studies employ an instrumental variable approach to estimate the returns to schooling (Card 2001). A valid instrumental variable must satisfy two main conditions: relevance and exogeneity.

---

[38] Three estimation caveats need to be highlighted. First, it is not clear how the relevant cognitive measure estimates are to be normalized. One possible solution is to express the effect of a unit change in cognitive ability on labor market outcomes in either units of percentage change in earnings or in standard deviation units of the distribution of earnings. Therefore, the coefficient of the cognitive measure in a log earnings specification provides the effect on the level of income, relative to mean incomes. Alternatively, the normalized regression coefficient provides the effect on income in units, based on the distribution. The first coefficient captures the effect of a unit increase in the cognitive score on earnings. Therefore, this is a reliable measure of the scarcity of cognitive skill. The second measure, which is normalized on the distribution of earnings, captures the importance of the distribution of cognitive skills as a determinant of the distribution of earnings. The square of the normalized regression coefficient captures the effect of the distribution of cognitive skills on the variance of the logarithm of earnings. These two estimates need not move in the same direction. We subsequently present estimates from a specification for the logarithm of the wage with respect to a normalized test score. Therefore, the percentage change in the wage is relative to a change in test scores by a standard deviation. Second, there are difficulties in estimation when we attempt to identify a time trend in the returns to cognitive performance, largely due to the fact that cognitive performance and educational attainment are highly sorted. A third problem stems from the timing of the calculation. The results depend, in part, on when we capture the wage returns to cognitive performance.

Relevance requires that the instrument be correlated with the number of years of schooling that an individual attains. Exogeneity requires that the instrument affect earnings only through the endogenous schooling variable. Previous studies explore a range of potential instruments, including changes in schooling laws (Harmon and Walker 1995), proximity to college (Card 1993), and birth quarters (Angrist and Krueger 1991).[39]

The *South African Schools Act* of 1996 requires students aged from 7 to 15 to complete schooling through the general education and training phase, or grade 9.[40] Because the requirement is based on when each student turns 15, each student becomes exempt from the requirement to complete grade 9 at different times. For example, an individual born in January will have had less education at the completion of the minimum school-leaving requirement than an individual born in May, following a September to June school-year schedule. Thus, this difference in schooling may lead to differences in earnings, which is arguably not correlated with ability. We use a person's quarter-of-birth indicator as a source of identifying variation in schooling completed. Therefore, we create quarter-of-birth dummy variables as instruments for schooling. A standard caveat to this instrumental variable approach is that the school-leaving age induces a sizable effect on when students leave school, so much so that the exogenous shock can be picked up by the proposed instruments. In other settings, some studies have documented that, due to seasonal food constraints, children's gestation during lean seasons may result in lower birth weights or cognitive ability and negatively influence their earnings (Almond and Mazumder 2011; Duncan et al. 2017; Lokshin and Radyakin 2012; Weber et al. 1998). Dorelien (2013) specifically examines for any pattern of seasonality of births in South Africa. The study finds that the pattern is "generally flat" (Dorelien 2013). The potential issue of any seasonality pattern due to seasonal food constraints will most likely only relate to the agricultural subsample in CAPS. The percentage of individuals in the CAPS sample who report their primary occupation in the agriculture, forestry, or fishing industry is approximately 0.5 percent. This implies that food seasonality is unlikely to constitute a major concern in the CAPS sample. However, we explore for seasonality of births, in particular, whether it is sensitivity to months related to the seasonality of food supply, following the methodology by Lam and Miron (1991) and Dorelien (2013).[41] In these additional analyses, we do not detect any evidence of seasonality of births in the sample.[42] This additional analysis further bolsters the validity of the quarter-of-birth instrument.

---

[39] Card (2001) notes that most IV estimates are larger than the corresponding OLS estimates (Card 2001). Several potential explanations could account for this discrepancy: the IV corrects measurement error, the IV induces new biases, or the IV induces increase in education along margins with high returns to more schooling (Imbens and Angrist 1994).
[40] The *South African Schools Act* (No. 84 of 1996) was passed in November 6, 1996 and came into effect on January 1, 1997. Specifically, Chapter 2, Section 3 (1) covers the educational amendment of the Act regarding compulsory education based on either age or grade completion.

Following Angrist and Krueger (1991), we specifically use data on a person's quarter of birth from the CAPS to estimate specification (4) by instrumenting $S_{j,i}$ with instruments $Q_{1i}$ to $Q_{3i}$, which denote one's birth quarters. We first regress the schooling variable on three quarter-of-birth dummy variables:

(4) $\quad S_i = \alpha_i + \delta_{1i}Q_{1i} + \delta_{2i}Q_{2i} + \delta_{3i}Q_{3i} + x_i'\pi_{1,x} + v_{1,i}.$

The second stage is based on specifications (1) to (2), as described earlier. In developing countries, the dropout rate before the minimum school-leaving age is likely to be measured more imprecisely than in developed countries. In addition, the measurement error in the self-reported schooling variable is likely to be extremely high. Therefore, a 2SLS analysis, using quarter of birth as an instrument for one's schooling and using data from developing countries, is challenging. Our analysis uses two sources of data: the Wave 5 CAPS and Panels 1, 3, 4, and 5 of the CAPS.

---

[41] Using data from the CAPS on the month of birth and year of birth for individuals in the CAPS sample, we calculated the *monthly birth amplitude* (the percentage deviation above the expected monthly number of births based on a uniform distribution of births within a year). For example, a monthly amplitude of 110 with an index at 100 means that there are 10 percent more births in that month in comparison with the annual monthly mean. Therefore, amplitude will demonstrate whether there is a birth peak or trough in a particular month. We regress the monthly amplitude on monthly dummies for each month; we do not detect any evidence of statistically significant monthly patterns.
[42] In results not reported here, we find no empirical evidence that the seasonality of birth influences any proxies for human capital investments.

## V. Results: Returns to Schooling and Bounding the Ability Bias

### A. Ordinary Least Squares Estimates of the Mincer's Earnings Model

*Cape Area Panel Study Sample.* Tables 3, 4, and 5 present the OLS estimates based on specifications (1) to (3), discussed in section IV.[43]

Table 3, column (1), presents the OLS estimates of the Mincer equation without measures of cognitive ability. Column (2) includes the aggregate cognitive ability index. Column (3) includes the distinct ability domain measures for the two ability domains in the CAPS, with one index for literacy and one for numeracy.[44] The coefficient estimate, reported in column (1), of the schooling variable shows that, for each additional year of schooling that is acquired, an individual's earnings increase by 14 percent. Once we account for the aggregate ability measure, the schooling coefficient, reported in column (2) and column (3), drops by two to three percent. The drop in the schooling coefficient estimate implies the presence of large ability bias in the OLS estimation of the return to schooling. This pattern of the estimated effect size for the schooling variable (i.e., approximately 11 percent of increased earnings associated with each additional year of schooling) is considerably smaller than previous empirical estimates of the returns to schooling in South Africa (Salisbury 2016). Table 3 also shows a substantial earnings gap between blacks/coloured and white. This earnings gap between is reduced, by approximately a third, once the ability measures are included.[45] In contrast, the gender earnings gap is not affected after the inclusion of the ability measures.

[Table 3 about here]

Table 4 reports the results based on the CAPS panel dataset, which is based on Waves 1, 3, 4, and 5. In Table 4, column (1), we report the OLS estimates without including an ability measure. The estimated effect size for the continuous schooling variable is a 16 percent increase in earnings for each additional year of schooling. Table 4, column (2), includes the aggregate ability measure. Table 4, column (3), presents estimates based on the separate cognitive domain measures; the effect size in both columns is a 14 percent increase, which is associated with an additional year of schooling, holding all else constant. The estimated coefficient of the returns to schooling based on the panel survey is only slightly larger compared with the estimated coefficient based only on Wave 5, once we control for ability measures.

---

[43] The reported standard errors are corrected for cluster correlation, following Cameron et al. (2008) and Esarey and Menger (2018).
[44] We also examine whether the raw proxies for ability measures are influenced by a schooling gradient at the time of the cognitive evaluations in Wave 1. We revisit the issue in detail in the sections covering additional robustness checks. In this additional robustness check, we reconstruct the ability proxy measures following the methodological approach in (Cascio and Lewis 2006; Dahl and Lochner 2012). Based on the methodology in these studies, we find no evidence that the reconstructed cognitive measures are causally dependent on differences in schooling levels. The results based on the reconstructed ability measure remain consistent in both magnitude and statistical significance with the results reported in this section.
[45] This result mirrors a similar finding found in Neal and Johnson (1996), a study that examines the U.S. racial wage gap using data from the National Longitudinal Survey of Youth.

[Table 4 about here]

*HAALSI Sample.* Table 5 reports the estimates from the OLS estimation using the HAALSI sample. Table 5 presents specification (1), the standard Mincer specification without an ability proxy, whereas specification (2) and specification (3) include the aggregate ability index (see column 2) or only the distinct cognitive domain measures (see column 3). Similar to the results presented in Table 3, the empirical estimates of the returns to schooling presented in Table 5 are strikingly similar.[46] The estimate implies approximately 10 percent higher earnings associated with each additional year of schooling acquired. The difference between columns (1) and (2)—the implied ability bias associated with this rural sample—is approximately one percent.

[Table 5 about here]

Furthermore, we break the ability score into specific cognitive domains, and we account for each domain in the Mincer specification. The only two measures of cognitive domains that are statistically significant are memory and orientation, with effect sizes of approximately 10 and 7 percent, respectively. Therefore, it is likely that the omission of some measures of these two cognitive domains (i.e., the measures for memory and orientation) accounts for the ability bias in the naïve OLS estimation. We cannot reject the null hypothesis that other cognitive domains of numeracy and attention are significant predictors of a person's earnings.

---

[46] Similar to the analysis based on the CAPS sample, the inclusion of the ability measures does not affect the gender earnings gap.

## B. Instrumental Variable Estimates of the Returns to Schooling

*Cape Area Panel Study Sample.* Next, we instrument for a person's schooling using data on their quarter of birth. Tables 6 and 7 present the 2SLS estimates.[47] Because the CAPS sample is already based on adolescents, the sample is restricted to adolescents who are subject to the compulsory school law. As noted previously, the exclusion restriction for the validity of Q1 to Q3 as instruments require no direct relationship between the quarter of birth and earnings. Thus, birth timing should not have a direct effect on a person's wages, but rather should affect wages only through the relationship with completed schooling induced by compulsory education laws. Table 6 presents estimates of the first stage, in which education is regressed on the quarter-of-birth instruments. Table 6 reports how the proposed instruments (Q1–Q3) that are used as a source of identifying variation for schooling ($S_{ji}$) influence a person's schooling decisions.[48] Although the results in Table 6 pass the standard F-test for a strong first stage, not all quarter-of-birth binary variables are statistically significant predictors of a person's educational attainment. Table 7 reports the second stage of the 2SLS estimation of the returns to education.[49]

[Table 6 about here]   [Table 7 about here]

Due to the extremely small sample size (401 observations), the second stage estimates are insignificant but are very close to the 10 percent level. The imprecisely estimated effect size on the schooling variable, however, does provide some information. Most of the IV-based estimation shows higher estimates of the returns to schooling compared with the OLS estimates. The effect size is approximately 28 percent, which is associated with an additional year of schooling attained, although the IV estimates are not significant at the 10 percent level.

To assess the impact of educational attainment and the cognitive measures, we also use an additional instrumental variable based on the CAPS data. Specifically, we use school fees in the CAPS as an instrument for a person's educational attainment. Factors from the supply side (related to schooling) are an obvious source of identifying variation that can be used for estimating demand-side parameters. Assuming that school fees are independent of taste and ability factors, we can use this variable as a source of exogenous variation in educational attainment[50] and in an auxiliary IV estimation scheme. In the U.S. context, empirical papers have previously used school fees as an instrument for schooling (Card 1999; Kane and Rouse 1993). In South Africa, enrollment at the secondary school level can be particularly sensitive to school fees (Fiske and Ladd 2003, 2004). Borkum (2012) uses data from South Africa and demonstrates that school enrollment is particularly sensitive to school fees at the secondary school level. For the CAPS sample, in Wave 1 and Wave 2, the survey questionnaire asked respondents about school fees that were paid.[51] We use this information to instrument for one's educational attainment with the total reported school fees.[52] We report the results for this additional 2SLS estimation in Table 8. Based on this instrument, although the effect size is slightly lower than the one that is based on the quarter-of-birth instrument, the pattern of the results remains consistent with the results reported in Table 7. Three explanations can generally account for the difference between the OLS and IV estimates. First, measurement error generally generates downward bias in the OLS estimates. Second, it could be that individuals of higher ability tend to get more education causing upwards bias in the OLS estimates; our proxies of ability could account for some but not for all variation in ability. Third, the IV estimation relies on identifying variation based on the so-called group of "compliers", which could have relatively high returns to education (Imbens and Angrist 1994; Angrist, Imbens and Rubin 1996; Imbens and Rubin 1997). In the context of our data, the difference between the OLS and IV estimates is likely due to the 2SLS estimates living off identifying variation from individuals with high marginal returns to schooling. In analyses not reported in the tables, we find that the complier population is comprised of individuals who are more likely to be very low-income, more likely to be female, more likely to be quite young (less than 20 years old), and are more likely to be non-white.

---

[47] We report clustered standard errors following Cameron et al. (2008) and Esarey and Menger (2018).
[48] We use a $_{ji}$ subscript (j denoting each survey wave) because the 2SLS estimation is based on all CAPS waves as a panel.
[49] Individuals younger than 15 years old are excluded, because they are not able to exercise the minimum school-leaving age exemption for compulsory schooling. The minimum school-leaving age is 15, which is typically the age of lower secondary school attainment. Therefore, the regression is restricted to a maximum schooling attainment of 10th grade. The distribution of schooling attainment in the CAPS sample has an average level of 7.59 among 15-year-olds, so it is unreasonable to assume that most 15-year-olds have attained the national standard of 9th grade by the age of 15. Instead, students in 7th and 8th grade are likely to be affected by the minimum school-leaving age.

---

[50] Grubb (1989), Rouse (1994, 1995, 1998), and Kane (1994, 1995) demonstrate that tuition fees and educational expenses can be an important determinant of educational attainment.
[51] Given the sample, the educational level for the nearest school refers to primary and secondary schools.
[52] School fees might not be a valid instrument in a setting in which families move to certain communities because of their characteristics (if these characteristics are related to the school fees) or if school placement is nonrandom and related to fees (Pitt et al. 1993; Rosenzweig and Wolpin 1986).

We report a summary of the main OLS results, based on the specifications above, in Table 9.

[Table 8 about here] [Table 9 about here]

C.     **Ordinary Least Square Estimates of Returns to Schooling by Schooling Levels**

Table 10 reports estimates of the returns to schooling regression by schooling tier for both the CAPS and HAALSI samples.

[Table 10 about here]

Table 10 reports the results with the reference category being individuals with no schooling or less than primary schooling. The estimated effect size for the various schooling tiers reports the highest returns to completing upper-secondary schooling. If a person completes upper-secondary schooling, his or her increase in earnings is approximately 140 percent higher than the earnings of a person with less than primary schooling or no schooling at all. A particularly striking finding is the magnitude of the coefficient estimates for the completion of upper secondary schooling among the rural HAALSI sample. The associated effect size for the completion of the upper-secondary schooling among the HAALSI sample is almost double that of the CAPS sample.

Table 10 reports the effect size for the aggregate standardized cognitive score for the HAALSI sample. In column (2), we note that a standard deviation increase in the aggregate cognitive score is comparable to the effect size associated with the returns to schooling if a person completes only primary schooling—relative to someone who completes no schooling at all. The effect size for the aggregate ability measure in the CAPS sample (i.e., the urban sample in South Africa) is slightly larger than the effect size for the aggregate ability measure based on HAALSI.

D.     **Effect Size Estimates on the Returns to Specific Cognitive Domains in the Rural and Urban Samples**

The results in Table 3, Table 5, and Table 10 also present estimates of the effect sizes of specific cognitive domain measurements for the labor market outcomes in the two samples. Of the four cognitive domains discussed in Section II.B, HAALSI captures three: attention, memory, and higher-order cognitive functions. HAALSI lacks a measure of inhibitory control. The CAPS captures only one major domain: the higher-order cognitive functions (captured by numeracy and literacy skills). HAALSI and CAPS overlap, as both study measures contain numeracy skills.

We examine the effect of each cognitive domain on individual earnings. Table 3 reports the effect sizes associated with the returns to specific cognitive skills in the CAPS (the urban sample). The reported results for numeracy are statistically significant at the 5 percent level. The effect size that is associated with numeracy is noticeably larger, denoting greater importance of numeracy skills for the earnings outcome among the CAPS sample. Although, from a causal standpoint, the results should be interpreted cautiously, the effect size associated with a standard deviation improvement in the numeracy test is of the same magnitude as the effect size associated with the continuous schooling variable.

Similarly, Table 5 reports the effect sizes associated with the returns to specific cognitive skills in HAALSI (the rural sample). Only the reported results for the domains of memory and orientation are statistically significant, and only the memory domain is at the 1 percent level. The effect size associated with memory skills is approximately equivalent to the magnitude of the effect size associated with the continuous schooling variable among the HAALSI respondents. Finally, Table 10 reports the effect sizes for both survey samples using dichotomous schooling level tiers. The pattern of the results reported in Table 3 and Table 5 remains consistent; only the numeracy skills strongly influence earnings in the CAPS sample. For the HAALSI sample, memory and orientation have the largest effect sizes, which are approximately the same. The effect of numeracy is measured in the HAALSI sample, although the effect size is not statistically significant.

Based on these two samples, the overarching pattern is that executive functioning is more important in the rural sample, whereas higher-order cognitive skills are more important in the urban sample. Executive functions include basic cognitive processes, such as attentional control, inhibitory control, working memory, and cognitive flexibility. As noted earlier, in the rural sample, memory and orientation are the only two domains that are statistically significant predictors of earnings. Both domains fall under executive functioning. Although numeracy is tested in both samples, it is a statistically significant predictor of earnings only in the urban sample (i.e., the CAPS data). There is limited information in the HAALSI sample, in addition to methodological differences between how the two surveys capture occupational and industry-level data among those who report being employed. As such, it is not possible to fully examine the extent to which differences in occupational distributions or sector capital levels between the two samples influence the returns to higher-order cognitive skills in the urban sample. It is likely that other competing explanations also play an important role—for example, differences between the demographic compositions of workers, age-specific effects, or urbanity and space differences that bid up returns to labor and cognitive domains due to demand differences across geographic space.

## VI. Robustness Checks

Next, we present a couple of robustness analyses to further investigate the magnitude of the 2SLS estimates of the returns to schooling. Specifically, we use an additional IV for a person's schooling. The second robustness check examines a potential concern related to the issue that the proxy measures for the cognitive skill domains could be affected by the level of schooling. We address this concern by reconstructing an alternative measure for cognitive abilities.

### A. Alternative Instrumental Variables

We further examine the validity of effect size estimates based on the 2SLS approach by using the idiosyncratic features of the South African educational system and in particular, the interplay between racial identity and access to education under apartheid. Specifically, we instrument an individual's educational attainment with his or her year of birth interacted with that person's population group status.[53] Appendix A reports the results based on this additional 2SLS approach (see Online Appendix Table A3). We find that all of the main effect estimates are in the range of the estimates we find using the main instrument (i.e., quarter of birth) reported in Tables 7-9.

---

[53] The assumption in this additional approach is that the effect of racial identity interacted with year of birth influences earnings only through individual educational attainment gradient, an assumption that is fundamentally untestable.

### B. Reconstructed Cognitive Evaluation Measures

Next, we attempt to address a potential issue in the cognitive scores in the CAPS. Following the pioneering work of Spearman (1927), previous research uses U.S. data and argues that age differentials may affect test scores because of differences in educational attainment (Cascio and Lewis 2006; Hansen et al. 2004; Keane and Wolpin 1997; Heckman and Vytlacil 2001) or maturity. We conduct additional statistical tests to test this hypothesis in our setting. We examine whether the literacy and numeracy evaluation measurements increase with age or whether individuals of the same age, but different educational gradients, exhibit higher literacy and numeracy measures. We report the results in Online Appendix B, Table B1 through to Table B5. We fail to detect evidence to support that the cognitive measures are affected by an educational gradient.[54] The pattern of the results remains consistent with the results reported in Table 3 through Table 10.

## VII. Discussion and Concluding Remarks

Numerous studies have investigated the monetary returns to education in developed countries. Most studies using data from developing countries that examine the relationship between years of education and wages are based on naïve OLS estimations. For sub-Saharan Africa, where rates of poverty are among the highest in the world (World Bank 2016), finding effective levers to boost educational outcomes are especially important. Therefore, the issue of the measured estimates for the returns to schooling is central for policymaking in developing countries. Furthermore, there is a growing consensus that cognitive abilities and personality traits can be important for labor market success (Deming 2017). Cognitive skills, especially those in specific cognitive domains, can translate to substantial improvements in job performance and boost productivity and earnings outcomes.

---

[54] The two exercises were intended to examine—in the spirit of Segal (2012)—whether there is empirical evidence regarding differences in cognitive measurements either for people of the same age but with different educational gradients or for people of the same educational attainment but of different maturity levels or ages (based on Segal 2012). Based on the F-test (for different age-groups or educational gradients), we fail to detect any evidence that the CAPS Wave 1 cognitive evaluations differed either by educational level or age at the time of the Wave 1 survey (i.e., maturity).

To this end, in this study, we examine the importance of cognitive domains for the returns to schooling in the context of sub-Saharan Africa using data from two labor market surveys in South Africa. The first survey, HAALSI, examines a rural population aged 40 and older. The second survey, CAPS, follows a metropolitan area population aged 14 to 22. Both collected cognitive evaluations via direct measurements of specific cognitive domains. Using these measures, we examine the influence of each cognitive domain on one's reported earnings. Furthermore, we compile the information from each cognitive domain into an aggregate ability index, and we examine how each cognitive measure influences one's earnings. We also examine how the aggregate ability measure affects the measured returns to attaining more schooling.

Using the HAALSI and CAPS samples, we find that the return for an additional year of schooling ranges from 9 to 14 percent. In the HAALSI sample, the estimate is approximately 9 percent of higher earnings associated with each additional year of schooling. Based on the CAPS sample, the estimated return was approximately 11 percent. We also find that, when accounting for ability measures in the Mincer earnings equation, the implied ability bias in the observational design estimates ranges from one to three percent. Although significant progress has been made in estimating the returns to schooling in high-income countries, for the most part, estimates of the educational returns in developing countries rely on OLS estimations. The choice of using OLS estimations is justified only if realized schooling and unobserved labor market ability are uncorrelated. Therefore, the validity of very high returns to schooling, especially those found in data for sub-Saharan countries—as reported by Psarachopolous and Patrinos (2004)—casts doubt about the true causal nature of these empirical estimates. If these estimates reflect true high-measured returns to schooling, it is unclear whether the rising premium for education reflects a higher return to formal schooling or a larger—and unaccounted for—ability bias in the estimated returns to schooling. This distinction and disentangling of the importance of each contributing factor is essential for human capital policy. If the higher education premium reflects the growing importance of unobserved ability factors that are acquired earlier in life, it may be better to divert resources from formal education to preschool education and childcare. Although we document evidence of positive ability bias using data from South

Africa, the magnitude of our estimates is lower than what has previously been found in high-income countries.[55]

Furthermore, we instrument for individual schooling levels using schooling fees with data from the CAPS. Using this approach and using the schooling fees as an instrument, we estimate the 2SLS effect size for the return to schooling: each additional year of schooling increases earnings by approximately 18-20 percent. This result implies that the 2SLS estimates are higher than the estimates based on the OLS approach. The finding that the IV estimates of the returns to education are higher than the OLS estimates is consistent with the summary by Card (1999) of the return to schooling in high-income countries. In our context, however, the estimators are significantly further apart.

We also find striking results regarding the importance of specific cognitive domains when we compare the urban and rural samples. The overarching pattern in the data demonstrates that executive functioning processes, such as attentional control and working memory, are more important for determining earnings for the rural sample. Conversely, higher-order cognitive skills have greater significance in the earnings for the urban sample. Although numeracy is a cognitive measure in both the rural and urban samples, it is only a statistically significant predictor of earnings in the urban sample.

Finally, we show evidence that demonstrates the importance of specific cognitive abilities. These findings introduce a caveat to the pervasive view of developing economies in the literature that general cognitive ability, as measured by test scores, is important for explaining personal labor market outcomes. Although cognitive skills certainly explain much of the variance in wages, specific cognitive domains (i.e., executive functioning in rural settings and higher-order cognitive skills in urban settings) play a more prominent role than general cognitive skills. These results have potential implications for understanding education and labor markets in developing countries. Although our research illuminates the important role of specific cognitive skills for determining a person's earnings, from a policy standpoint, much more must be learned about potential policy interventions—in addition to their cost and timing over one's lifecycle—that could effectively influence these important cognitive domains.

---

[55] See Kaymak (2009) for U.S. estimates.

# FIGURES AND TABLES

TABLE 1. EDUCATIONAL SYSTEM BY YEARS OF SCHOOLING IN SOUTH AFRICA

| Years of Schooling | Grades[a] | Diplomas/Degrees in South Africa |
|---|---|---|
| 7 | Grade R plus Grades 1 to 7 | Primary level |
| 9 | Grades 8-9 | Lower Secondary level |
| 12 | Grades 10-12 | Upper Secondary level |
| 13-16 | NTC I, II, III/Diploma/Certificate[b] | Technical Tertiary level[d] |
| 13-16 | Undergraduate Diploma/Undergraduate degree/Postgraduate degree or diploma[c] | University Tertiary level[e] |

*Notes*: (a) Based on Lehohla (2017), UNESCO (2012), and Krige et al. (1994). (b) Based on d'Almaine, Manhire and Atteh (1997) and South African Qualifications Authority (2017). (c) Based on NAFSA (2016). (d) Technical Tertiary level schooling refers to studies in a vocational discipline. (e) University Tertiary level schooling refers to tertiary studies in an academic discipline.

TABLE 2. SUMMARY STATISTICS HAALSI AND CAPS WAVE 5

| | Metropolitan Cape Town[a] CAPS 2009 | Agincourt, South Africa HAALSI 2014 | QLFS[d] 2009 | QLFS[d] 2009 (Urban, Age 21-29) | QLFS[d] 2009 (Rural, Age >35) | QLFS[d] 2014 | QLFS[d] 2014 (Urban, Age 26-34) | QLFS[d] 2014 (Rural, Age>40) |
|---|---|---|---|---|---|---|---|---|
| | (1) | (2) | (3) | (4) | (5) | (6) | (7) | (8) |
| Gender (1 if male) | 0.45 | 0.46 | 0.49 | 0.51 | 0.51 | 0.49 | 0.51 | 0.40 |
| | (0.5) | (0.5) | (0.50) | (0.50) | (0.50) | (0.50) | (0.50) | (0.49) |
| Age (in years) | 24.88 | 54.51 | 27.37 | 25.04 | 29.92 | 28.49 | 29.92 | 56.89 |
| | (2.48) | (15.35) | (19.28) | (2.54) | (2.52) | (19.63) | (2.52) | (12.30) |
| Schooling (in years) | 10.76 | 3.68 | 7.02 | 11.02 | 11.30 | 7.48 | 11.30 | 5.68 |
| | (2.35) | (4.42) | (4.77) | (2.28) | (2.40) | (4.76) | (2.40) | (4.49) |
| Log Monthly Earnings[b] | 7.52 | 7.11 | 7.79[e] | 7.54[e] | 7.95[e] | 7.87[e] | 7.80[e] | 7.98[e] |
| | (0.64) | (1.39) | (1.27) | (1.21) | (1.32) | (1.19) | (1.14) | (1.30) |
| Experience[c] | 8.11 | 44.67 | 14.30 | 7.98 | 12.56 | 15.05 | 12.56 | 42.97 |
| | (3.21) | (17.37) | (16.82) | (3.28) | (3.39) | (17.18) | (3.39) | (13.20) |
| Married (1 if married) | 0.14 | 0.51 | 0.21 | 0.12 | 0.22 | 0.20 | 0.22 | 0.43 |
| | (0.35) | (0.5) | (0.41) | (0.33) | (0.42) | (0.40) | (0.42) | (0.49) |
| Black African (1 if yes) | 0.45 | | 0.79 | 0.75 | 0.77 | 0.80 | 0.77 | 0.96 |
| | (0.50) | | (0.41) | (0.43) | (0.42) | (0.40) | (0.42) | (0.21) |
| Coloured[f] (1 if yes)[f] | 0.42 | | 0.09 | 0.11 | 0.10 | 0.09 | 0.10 | 0.02 |
| | (0.49) | | (0.29) | (0.32) | (0.31) | (0.29) | (0.31) | (0.15) |
| White (1 if yes) | 0.13 | | 0.12 | 0.13 | 0.12 | 0.11 | 0.12 | 0.02 |
| | (0.33) | | (0.32) | (0.34) | (0.33) | (0.31) | (0.33) | (0.15) |
| Never attended school | 0.00 | 0.45 | 0.18 | 0.01 | 0.01 | 0.16 | 0.01 | 0.25 |
| | (0.00) | (0.50) | (0.38) | (0.07) | (0.08) | (0.37) | (0.08) | (0.43) |
| Below Primary | 0.04 | 0.28 | 0.24 | 0.04 | 0.03 | 0.21 | 0.03 | 0.29 |
| | (0.19) | (0.45) | (0.42) | (0.19) | (0.17) | (0.41) | (0.17) | (0.45) |
| Attained Primary Education Only | 0.13 | 0.10 | 0.12 | 0.07 | 0.05 | 0.11 | 0.05 | 0.17 |
| | (0.34) | (0.3) | (0.33) | (0.25) | (0.22) | (0.31) | (0.22) | (0.37) |
| Attained Lower Secondary Only | 0.40 | 0.07 | 0.23 | 0.35 | 0.35 | 0.25 | 0.35 | 0.17 |
| | (0.49) | (0.26) | (0.42) | (0.48) | (0.48) | (0.44) | (0.48) | (0.37) |
| Attained Upper Secondary Only | 0.27 | 0.05 | 0.16 | 0.41 | 0.38 | 0.18 | 0.38 | 0.08 |
| | (0.44) | (0.21) | (0.37) | (0.49) | (0.49) | (0.39) | (0.49) | (0.26) |
| Attained Some Tertiary | 0.16 | 0.04 | 0.07 | 0.12 | 0.18 | 0.08 | 0.18 | 0.05 |
| | (0.37) | (0.20) | (0.25) | (0.33) | (0.38) | (0.27) | (0.38) | (0.21) |
| Observations | 4,752 | 5,038 | 88,252 | 8,287 | 7,406 | 85,302 | 7,406 | 9,029 |

*Notes*: Earnings is calculated using 2002 constant ZARs. (a) CAPS summary statistics based on Wave 5, 2009 data. (b) Earnings are in Constant 2002 ZARs (c) Experience is calculated by taking the difference of one's age and one's schooling minus six years following Mincer (1974), Boissiere, Knight and Sabot (1985) and Lemieux (2006). Primary, Lower Secondary, and Upper Secondary education refers to individuals who completed only primary, lower secondary, and upper secondary schooling, respectively. (d) The Quarterly Labour Force Survey (QLFS) in South Africa uses the master sample frame which has been developed as a general-purpose household survey frame for labor market issues in South Africa. The survey's master sample is designed to be representative at provincial level and within provinces at metro/non-metro levels. (e) Earnings are based on data from the South Africa's Income Dynamics. (f) Coloured is the term for a person of mixed European ("white") and African ("black") or Asian ancestry, as officially defined by the South African government from 1950 to 1991 and by the current national statistical bureau of South Africa (called Statistics SA) (Statistics SA 2017). Standard deviations in parentheses.

TABLE 3. CAPS SURVEY (WAVE 5): RETURNS TO EDUCATION, OLS ESTIMATES

|  | Log of Monthly Earnings (ZAR) | | |
|---|---|---|---|
|  |  | With Ability[b] | With Specific Ability Domains[c] |
|  | (1) | (2) | (3) |
| Schooling (in years) | 0.14*** | 0.12*** | 0.11*** |
|  | (0.01) | (0.01) | (0.01) |
| Experience[a] | 0.05*** | 0.04** | 0.04** |
|  | (0.02) | (0.02) | (0.02) |
| Experience Squared | -0.00 | -0.00 | -0.00 |
|  | (0.00) | (0.00) | (0.00) |
| Male (=1 if male) | 0.19*** | 0.18*** | 0.17*** |
|  | (0.03) | (0.03) | (0.03) |
| Black (=1 if black) | -0.60*** | -0.48*** | -0.44*** |
|  | (0.10) | (0.11) | (0.11) |
| Coloured[d] (=1 if coloured) | -0.15 | -0.08 | -0.05 |
|  | (0.10) | (0.10) | (0.10) |
| Aggregate Ability |  | 0.10*** |  |
|  |  | (0.02) |  |
| Literacy Component |  |  | 0.01 |
|  |  |  | (0.02) |
| Numeracy Component |  |  | 0.11*** |
|  |  |  | (0.02) |
| Constant | 5.95*** | 6.14*** | 6.15*** |
|  | (0.16) | (0.16) | (0.16) |
| $R^2$ statistic | 0.27 | 0.28 | 0.28 |
| Observations | 1,973 | 1,973 | 1,973 |

*Notes*: Cape Area Panel Study. Earnings is in Constant 2002 ZARs. Exchange rate from 2002 is 10.539 USD-ZAR. [a]Experience is calculated by taking the difference of one's age and one's schooling minus six years following Mincer (1974), Boissiere, Knight and Sabot (1985) and Lemieux (2006). (b) Aggregate ability is calculated using principal component analysis on the Literacy and Numeracy Evaluation survey questions. (c) Components of Literacy and Numeracy Evaluation are calculated using principal component analysis on two sets of survey questions: literacy-related questions and numeracy-related questions. (d) Coloured is the term for a person of mixed European ("white") and African ("black") or Asian ancestry, as officially defined by the South African government from 1950 to 1991 and by the current national statistical bureau of South Africa (called Statistics SA) (Statistics SA 2017). Standard errors are clustered at the neighborhood level.
*** Significant at the 1 percent level.
** Significant at the 5 percent level.
* Significant at the 10 percent level

TABLE 4. CAPS SURVEY (PANEL 1-3-4-5): RETURNS TO EDUCATION, OLS ESTIMATES

| | Log of Monthly Earnings (ZAR) | | |
|---|---|---|---|
| | | With Ability[b] | With Specific Ability Domains[c] |
| | (1) | (2) | (3) |
| Schooling (in years) | 0.16*** | 0.14*** | 0.14*** |
| | (0.02) | (0.02) | (0.02) |
| Experience[a] | 0.14*** | 0.16*** | 0.15*** |
| | (0.03) | (0.03) | (0.03) |
| Experience Squared | -0.01*** | -0.01*** | -0.01*** |
| | (0.00) | (0.00) | (0.00) |
| Gender (=1 if male) | 0.24*** | 0.24*** | 0.23*** |
| | (0.08) | (0.07) | (0.07) |
| Black (=1 if black) | -0.56*** | -0.40** | -0.36** |
| | (0.17) | (0.18) | (0.18) |
| Coloured[d] (=1 if coloured) | -0.22 | -0.14 | -0.09 |
| | (0.14) | (0.14) | (0.15) |
| Aggregate Ability | | 0.13*** | |
| | | (0.04) | |
| Literacy Component | | | -0.02 |
| | | | (0.05) |
| Numeracy Component | | | 0.15*** |
| | | | (0.05) |
| Wave 3 Dummy | 0.49*** | 0.49*** | 0.49*** |
| | (0.07) | (0.07) | (0.07) |
| Wave 4 Dummy | 0.55*** | 0.55*** | 0.55*** |
| | (0.07) | (0.07) | (0.07) |
| Wave 5 Dummy | 0.50*** | 0.50*** | 0.51*** |
| | (0.10) | (0.10) | (0.10) |
| Constant | 4.97*** | 5.06*** | 5.03*** |
| | (0.24) | (0.24) | (0.24) |
| $R^2$ statistic | 0.41 | 0.43 | 0.43 |
| Observations | 940 | 940 | 940 |

*Notes*: Cape Area Panel Study. Earnings is in Constant 2002 ZARs. Exchange rate from 2002 is 10.539 USD-ZAR. (a) Experience is calculated by taking the difference of one's age and one's schooling minus six years following Mincer (1974), Boissiere, Knight and Sabot (1985) and Lemieux (2006). (b) Aggregate ability is calculated using principal component analysis on the Literacy and Numeracy Evaluation survey questions. (c) Components of Literacy and Numeracy Evaluation are calculated using principal component analysis on two sets of survey questions: literacy-related questions and numeracy-related questions. Wave dummies work as follows: wave 3 dummy=1 if observation is from wave 3. wave 4 dummy=1 if observation is from wave 4. wave 5 dummy=1 if observation is from wave 5. (d) Coloured is the term for a person of mixed European ("white") and African ("black") or Asian ancestry, as officially defined by the South African government from 1950 to 1991 and by the current national statistical bureau of South Africa (called Statistics SA) (Statistics SA 2017). Standard errors are clustered at the neighborhood level.
*** Significant at the 1 percent level.
** Significant at the 5 percent level.
* Significant at the 10 percent level

TABLE 5. HAALSI SURVEY: RETURNS TO EDUCATION, OLS ESTIMATES

|  | Log of Monthly Earnings (ZAR) | | |
|---|---|---|---|
|  |  | With Ability[b] | With Specific Ability Domains[c] |
|  | (1) | (2) | (3) |
| Schooling (in years) | 0.10*** | 0.09*** | 0.09*** |
|  | (0.01) | (0.01) | (0.01) |
| Experience[c] | 0.03*** | 0.03*** | 0.03*** |
|  | (0.01) | (0.01) | (0.01) |
| Experience Squared | -0.00** | -0.00** | -0.00** |
|  | (0.00) | (0.00) | (0.00) |
| Gender (=1 if male) | 0.92*** | 0.90*** | 0.91*** |
|  | (0.05) | (0.05) | (0.05) |
| Aggregate Ability |  | 0.14*** |  |
|  |  | (0.03) |  |
| Memory Component |  |  | 0.10*** |
|  |  |  | (0.03) |
| Numeracy Component |  |  | -0.00 |
|  |  |  | (0.04) |
| Attention Component |  |  | 0.01 |
|  |  |  | (0.03) |
| Orientation Component |  |  | 0.07* |
|  |  |  | (0.04) |
| Constant | 5.16*** | 5.17*** | 5.12*** |
|  | (0.20) | (0.20) | (0.20) |
| $R^2$ statistic | 0.25 | 0.25 | 0.25 |
| Observations | 2,252 | 2,252 | 2,252 |

*Notes*: 2014-2015 Health and Aging in Africa: A Longitudinal Study of an INDEPTH Community in South Africa (HAALSI). Earnings data is based on last month's earnings. Earnings is in Constant 2002 ZARs. Exchange rate from 2002 is 10.539 USD-ZAR. Log earnings is calculated using earnings data in Rand. (a) Experience is calculated by taking the difference of one's age and one's schooling minus six years following Mincer (1974), Boissiere, Knight and Sabot (1985) and Lemieux (2006). (b) Aggregate ability is calculated using principal component analysis on the Cognition section of HAALSI. (c) Components of HAALSI Cognition section are calculated using principal component analysis on four sets of survey questions: memory-related questions, numeracy-related questions, attention-related questions, and orientation-related questions.
*** Significant at the 1 percent level.
** Significant at the 5 percent level.
* Significant at the 10 percent level

TABLE 6. CAPS SURVEY (WAVES 1-3-4-5): RETURNS TO EDUCATION, FIRST STAGE ESTIMATES WITH RESTRICTIONS. OLS ESTIMATES.

| CAPS WAVE 1-3-4-5 | | | | | |
|---|---|---|---|---|---|
| | | Quarter-of-birth effect | | | F-stat |
| | Mean | I | II | III | (P-value) |
| Years of Education | 8.04 | -0.092 | -0.04 | 0.35 | 3045.36 |
| | | (0.29) | (0.31) | (0.34) | (0.00) |
| Observations | 401 | | | | |

*Notes*: Data source: Cape Area Panel Study (2002-2009). The sample includes employed males and females of Metropolitan Cape Town aged 13 to 30 years. Standard errors are clustered at the neighborhood level. The sample is restricted to individuals who are affected by South Africa's Schools Act (No. 84) of 1996 and the restrictions are categorized by age or both age and grade. The age restriction is a minimum age of 15 to capture the school-leaving age. The grade restriction is a maximum schooling attainment of 9th grade, or the equivalent of school-leaving attainment. This eliminates learners who continued to study past the minimum school-leaving age, and are unlikely to be affected by the instrument as per Angrist and Krueger (1991).
*** Significant at the 1 percent level.
** Significant at the 5 percent level.
* Significant at the 10 percent level.

TABLE 7. CAPS SURVEY (WAVES 1-3-4-5): RETURNS TO EDUCATION, SECOND-STAGE ESTIMATION. 2SLS ESTIMATES

| Variables | Log of Monthly Earnings (ZAR) | | |
|---|---|---|---|
| | 2SLS Estimates[d] | | |
| | (1) | (2) | (3) |
| Schooling | 0.28 | 0.28 | 0.28 |
| | (0.19) | (0.18) | (0.19) |
| Experience[a] | 0.27*** | 0.27*** | 0.28*** |
| | (0.03) | (0.03) | (0.03) |
| Experience Squared | -0.01*** | -0.01*** | -0.01*** |
| | (0.00) | (0.00) | (0.00) |
| Gender (=1 if male) | 0.45*** | 0.46*** | 0.45*** |
| | (0.10) | (0.10) | (0.09) |
| Black (=1 if black) | 0.1 | 0.19 | 0.19 |
| | (0.59) | (0.59) | (0.59) |
| Coloured[e] (=1 if coloured) | 0.57 | 0.63 | 0.64 |
| | (0.58) | (0.58) | (0.58) |
| Ability[b] | | 0.06 | |
| | | (0.06) | |
| Literacy Component[c] | | | -0.01 |
| | | | (0.06) |
| Numeracy Component[c] | | | 0.08 |
| | | | (0.05) |
| Constant | 2.67 | 2.65 | 2.67 |
| | (1.96) | (1.81) | (1.85) |
| F-statistic | 3045.36 | 3045.36 | 3045.36 |
| $R^2$ statistic | 0.37 | 0.38 | 0.38 |
| Observations | 401 | 401 | 401 |

*Notes*: Cape Area Panel Study. Earnings is in Constant 2002 ZARs. Exchange rate from 2002 is 10.539 USD-ZAR. (a) Experience is calculated by taking the difference of one's age and one's schooling minus six years following Mincer (1974), Boissiere, Knight and Sabot (1985) and Lemieux (2006). (b) Aggregate ability is calculated using principal component analysis on the Literacy and Numeracy Evaluation survey questions. (c) Components of Literacy and Numeracy Evaluation are calculated using principal component analysis on two sets of survey questions: literacy-related questions and numeracy-related questions. (d) Restrictions are categorized by age or both age and grade. The age restriction is a minimum age of 15 to capture the school-leaving age. The grade restriction is a maximum schooling attainment of 9th grade, or the equivalent of school-leaving attainment. This eliminates learners who continued to study past the minimum school-leaving age, and are unlikely to be affected by the instrument as per Angrist and Krueger (1991). (e) Coloured is the term for a person of mixed European ("white") and African ("black") or Asian ancestry, as officially defined by the South African government from 1950 to 1991 and by the current national statistical bureau of South Africa (called Statistics SA) (Statistics SA 2017). Standard errors are clustered at the neighborhood level.

*** Significant at the 1 percent level.
** Significant at the 5 percent level.
* Significant at the 10 percent level.

TABLE 8. CAPS SURVEY: RETURNS TO EDUCATION, SECOND-STAGE ESTIMATION. BASED ON INSTRUMENTAL VARIABLES ESTIMATION WITH INSTRUMENT SCHOOL TOTAL FEES, 2SLS ESTIMATES

| Variables | Log of Monthly Earnings (ZAR) | | |
|---|---|---|---|
| | 2SLS Estimates | | |
| | (1) | (2) | (3) |
| Schooling | 0.20*** | 0.18*** | 0.18** |
| | (0.06) | (0.07) | (0.07) |
| Experience[a] | 0.27*** | 0.27*** | 0.27*** |
| | (0.02) | (0.02) | (0.02) |
| Experience Squared | -0.01*** | -0.01*** | -0.01*** |
| | (0.00) | (0.00) | (0.00) |
| Male (=1 if male) | 0.29*** | 0.29*** | 0.28*** |
| | (0.09) | (0.08) | (0.08) |
| Black (=1 if black) | -0.76*** | -0.61*** | -0.61*** |
| | (0.19) | (0.17) | (0.17) |
| Coloured[d] (=1 if coloured) | -0.35** | -0.28* | -0.26* |
| | (0.16) | (0.14) | (0.15) |
| Ability[b] | | 0.13 | |
| | | (0.08) | |
| Literacy Component[c] | | | 0.02 |
| | | | (0.06) |
| Numeracy Component[c] | | | 0.12** |
| | | | (0.06) |
| Constant | 4.46*** | 4.57*** | 4.61*** |
| | (0.82) | (0.84) | (0.83) |
| F-statistic | 4352.08 | 4352.08 | 4352.08 |
| $R^2$ statistic | 0.40 | 0.40 | 0.41 |
| Observations | 892 | 892 | 892 |

*Notes*: Cape Area Panel Study. Earnings is in Constant 2002 ZARs. Exchange rate from 2002 is 10.539 USD-ZAR. (a) Experience is calculated by taking the difference of one's age and one's schooling minus six years following Mincer (1974), Boissiere, Knight and Sabot (1985) and Lemieux (2006). (b) Aggregate ability is calculated using principal component analysis on the Literacy and Numeracy Evaluation survey questions. (c) Components of Literacy and Numeracy Evaluation are calculated using principal component analysis on two sets of survey questions: literacy-related questions and numeracy-related questions. Restrictions are categorized by age or both age and grade. The age restriction is a minimum age of 15 to capture the school-leaving age. The grade restriction is a maximum schooling attainment of 9[th] grade, or the equivalent of school-leaving attainment. This eliminates learners who continued to study past the minimum school-leaving age, and are unlikely to be affected by the instrument as per Angrist and Krueger (1991). (d) Coloured is the term for a person of mixed European ("white") and African ("black") or Asian ancestry, as officially defined by the South African government from 1950 to 1991 and by the current national statistical bureau of South Africa (called Statistics SA) (Statistics SA 2017). Standard errors are clustered at the neighborhood level.
*** Significant at the 1 percent level.
** Significant at the 5 percent level.
* Significant at the 10 percent level.

TABLE 9: SUMMARY OF SCHOOLING AND COGNITION RETURNS (CAPS AND HAALSI)

| Dependent variable: | | | | | | | Log of Monthly Earnings (ZAR) | | | | | | | | |
|---|---|---|---|---|---|---|---|---|---|---|---|---|---|---|---|
| | OLS Estimation | | | | | | 2SLS Estimation | | | | | | | | |
| Sample: | CAPS | | | HAALSI | | | CAPS | | | | | | | | |
| | | | | | | | Quarter of birth | | | School Fees | | | Birth year interacted with non-white status | | |
| | (1) | (2) | (3) | (4) | (5) | (6) | (7) | (8) | (9) | (10) | (11) | (12) | (13) | (14) | (15) |
| Schooling | 0.16*** (0.02) | 0.14*** (0.02) | 0.14*** (0.02) | 0.10*** (0.01) | 0.09*** (0.01) | 0.09*** (0.01) | 0.28 (0.19) | 0.28 (0.18) | 0.28 (0.19) | 0.20*** (0.06) | 0.18*** (0.07) | 0.18** (0.07) | 0.23 (0.31) | 0.25 (0.26) | 0.27 (0.25) |
| Aggregate Ability[b] | | 0.13*** (0.04) | | | 0.14*** (0.03) | | | 0.06 (0.06) | | | 0.13 | | | 0.07 (0.24) | |
| Literacy[c] | | | -0.02 (0.05) | | | | | | -0.01 (0.06) | | | 0.02 (0.06) | | | -0.03 (0.15) |
| Numeracy[c] | | | 0.15*** (0.05) | | | -0.00 (0.04) | | | 0.08 (0.05) | | | 0.12** (0.06) | | | 0.07 (0.13) |
| Memory[d] | | | | | | 0.10*** (0.03) | | | | | | | | | |
| Attention[d] | | | | | | 0.01 (0.03) | | | | | | | | | |
| Orientation[d] | | | | | | 0.07* (0.04) | | | | | | | | | |
| F-statistic | | | | | | | 3045.36 | 3045.36 | 3045.36 | 4352.08 | 4352.08 | 4352.08 | 1580.14 | 1580.14 | 1580.14 |
| $R^2$ statistic | 0.41 | 0.43 | 0.43 | 0.25 | 0.25 | 0.25 | 0.37 | 0.38 | 0.38 | 0.40 | 0.40 | 0.40 | 0.40 | 0.40 | 0.41 |
| Observations | 940 | 940 | 940 | 2,252 | 2,252 | 2,252 | 401 | 401 | 401 | 892 | 892 | 892 | 940 | 940 | 940 |

*Notes:* Earnings in Constant 2002 ZARs. Exchange rate from 2002 is 10.539 USD-ZAR. (a) Experience is calculated by taking the difference of one's age and one's schooling minus six years following Mincer (1974), Boissiere, Knight and Sabot (1985) and Lemieux (2006). (b) Aggregate ability is calculated using principal component analysis on the literacy and numeracy evaluation domains (for CAPS) and the memory, numeracy, attention, and orientation domains (for HAALSI). (c) Components of Literacy and Numeracy Evaluation are calculated using principal component analysis on two sets of survey questions: literacy-related questions and numeracy-related questions. Restrictions are categorized by age or both age and grade. The age restriction is a minimum age of 15 to capture the school-leaving age. The grade restriction is a maximum schooling attainment of 9th grade, or the equivalent of school-leaving attainment. This eliminates learners who continued to study past the minimum school-leaving age, and are unlikely to be affected by the instrument as per Angrist and Krueger (1991). (d) HAALSI comprises four cognition domains: memory, numeracy, attention, and orientation. Standard errors are clustered at the neighborhood level.
*** Significant at the 1 percent level.
** Significant at the 5 percent level.
* Significant at the 10 percent level.

TABLE 10. HAALSI AND CAPS WAVE 5 SURVEYS: RETURNS TO SCHOOLING BY EDUCATIONAL LEVEL, OLS

| Variables | Log of Monthly Earnings (ZAR) | | | | | |
|---|---|---|---|---|---|---|
| | HAALSI | | | CAPS Wave 5 | | |
| | | With Ability[b] | With Specific Ability Domains[c] | | With Ability[d] | With Specific Ability Domains[e] |
| | (1) | (2) | (3) | (4) | (5) | (6) |
| Primary (7-9) | 0.35*** | 0.29*** | 0.29*** | 0.13* | 0.07 | 0.08 |
| | (0.07) | (0.07) | (0.07) | (0.07) | (0.07) | (0.08) |
| Lower Secondary (10-11) | 0.58*** | 0.50*** | 0.50*** | 0.33*** | 0.21*** | 0.21*** |
| | (0.09) | (0.09) | (0.09) | (0.08) | (0.08) | (0.08) |
| Upper Secondary (12 or more) | 1.40*** | 1.29*** | 1.28*** | 0.77*** | 0.59*** | 0.58*** |
| | (0.08) | (0.08) | (0.08) | (0.08) | (0.08) | (0.09) |
| Experience[a] | 0.04*** | 0.04*** | 0.04*** | 0.03 | 0.02 | 0.03 |
| | (0.01) | (0.01) | (0.01) | (0.02) | (0.02) | (0.02) |
| Experience Squared | -0.00*** | -0.00*** | -0.00*** | 0 | 0 | 0 |
| | (0.00) | (0.00) | (0.00) | (0.00) | (0.00) | (0.00) |
| Male (=1 if male) | 0.96*** | 0.92*** | 0.92*** | 0.17*** | 0.16*** | 0.15*** |
| | (0.05) | (0.05) | (0.05) | (0.03) | (0.03) | (0.03) |
| Black (=1 if black) | | | | -0.67*** | -0.50*** | -0.46*** |
| | | | | (0.11) | (0.11) | (0.11) |
| Coloured[f] (=1 if coloured) | | | | -0.24** | -0.15 | -0.11 |
| | | | | (0.11) | (0.11) | (0.11) |
| Aggregate Ability | | 0.19*** | | | 0.12*** | |
| | | (0.03) | | | (0.02) | |
| Literacy Component | | | | | | 0.01 |
| | | | | | | (0.02) |
| Numeracy Component | | | 0.04 | | | 0.13*** |
| | | | (0.04) | | | (0.02) |
| Memory Component | | | 0.09*** | | | |
| | | | (0.03) | | | |
| Attention Component | | | 0.01 | | | |
| | | | (0.03) | | | |
| Orientation Component | | | 0.10*** | | | |
| | | | (0.04) | | | |
| Constant | 5.31*** | 5.26*** | 5.24*** | 7.19*** | 7.23*** | 7.20*** |
| | (0.20) | (0.19) | (0.20) | (0.13) | (0.13) | (0.14) |
| $R^2$ statistic | 0.25 | 0.26 | 0.26 | 0.25 | 0.27 | 0.27 |
| Observations | 2,252 | 2,252 | 2,252 | 1,973 | 1,973 | 1,973 |

*Notes*: Cape Area Panel Study and HAALSI. Earnings is in Constant 2002 ZARs. Exchange rate from 2002 is 10.539 USD-ZAR. (a) Experience is calculated by taking the difference of one's age and one's schooling minus six years following Mincer (1974), Boissiere, Knight and Sabot (1985) and Lemieux (2006). (b) Aggregate ability is calculated using principal component analysis on the Cognition section of the HAALSI survey. (c) Components are calculated using principal component analysis on four sets of survey questions pertaining to each cognitive domain. (d) Aggregate ability is calculated using principal component analysis on the Literacy and Numeracy Evaluation survey questions. (e) Components of Literacy and Numeracy Evaluation are calculated using principal component analysis on two sets of survey questions: literacy-related questions and numeracy-related questions. (f) Coloured is the term for a person of mixed European ("white") and African ("black") or Asian ancestry, as officially defined by the South African government from 1950 to 1991 and by the current national statistical bureau of South Africa (called Statistics SA 2017). Standard errors in column 4-6 are clustered at the neighborhood level.
\*\*\* Significant at the 1 percent level.
\*\* Significant at the 5 percent level.
\* Significant at the 10 percent level.

# ONLINE APPENDIX A

TABLE A1. CAPS PANEL SUMMARY STATISTICS

| | Wave 1 | Wave 2 | Wave 3 | Wave 4 | Wave 5 |
|---|---|---|---|---|---|
| Variable | (1) | (2) | (3) | (4) | (5) |
| Gender (1 if male) | 0.45 | 0.46 | 0.46 | 0.45 | 0.45 |
| | (0.50) | (0.50) | (0.50) | (0.50) | (0.50) |
| Married (1 if married) | 0.01 | | | 0.08 | 0.14 |
| | (0.10) | | | (0.27) | (0.35) |
| Black African (1 if yes) | 0.45 | 0.45 | 0.45 | 0.45 | 0.45 |
| | (0.50) | (0.50) | (0.50) | (0.50) | (0.50) |
| Coloured[b] (1 if yes) | 0.42 | 0.42 | 0.42 | 0.42 | 0.42 |
| | (0.49) | (0.49) | (0.49) | (0.49) | (0.49) |
| White (1 is yes) | 0.13 | 0.13 | 0.13 | 0.13 | 0.13 |
| | (0.33) | (0.33) | (0.33) | (0.33) | (0.33) |
| Schooling (in years) | 9.25 | 10.1 | 10.4 | 10.57 | 10.76 |
| | (2.15) | (2.14) | (2.20) | (2.24) | (2.35) |
| Experience[a] | 2.64 | 3.78 | 4.48 | 5.31 | 8.11 |
| | (2.11) | (2.60) | (2.84) | (2.99) | (3.21) |
| Earnings | 1167.74 | 9981.95 | 1924.68 | 2140.49 | 2189.63 |
| | (1318.45) | (26642.15) | (1890.57) | (1780.96) | (1614.5) |
| Age | 17.88 | 19.20 | 20.59 | 21.51 | 24.48 |
| | (2.48) | (2.51) | (2.52) | (2.56) | (2.58) |
| Observations | 4,752 | 3,927 | 3,531 | 3,439 | 2,915 |

*Notes*: Data source is Cape Area Panel Survey. The mean and standard deviation of each variable, with the standard deviation shown in parenthesis. Earnings is calculated using 2002 constant ZARs. (a) Experience is calculated by taking the difference of one's age and one's schooling minus six years following Mincer (1974), Boissiere, Knight and Sabot (1985) and Lemieux (2006). (b) Coloured is the term for a person of mixed European ("white") and African ("black") or Asian ancestry, as officially defined by the South African government from 1950 to 1991 and by the current national statistical bureau of South Africa (called Statistics SA) (Statistics SA 2017). Marital status for Wave 2 was not recorded in the original dataset, so it is marked as missing.

TABLE A2. COGNITIVE DOMAINS IN THE CAPS AND HAALSI SURVEYS

| Cognitive Domain | Description | CAPS | HAALSI |
|---|---|---|---|
| Literacy | language and literacy | X | |
| Numeracy | numeracy skills | X | X |
| Memory | word memorization | | X |
| Orientation | current event and date recognition | | X |
| Attention | self-rated attention and response to auditory stimuli | | X |

*Notes:* Data from HAALSI and CAPS. HAALSI cognitive domain components are defined using four sets of survey questions pertaining to each cognitive domain. CAPS cognitive domains are defined by the Literacy and Numeracy Evaluation (LNE) module of the CAPS survey. The LNE consists of two sets of survey questions: literacy-related questions and numeracy-related questions.

TABLE A3. CAPS SURVEY: RETURNS TO EDUCATION, SECOND-STAGE ESTIMATION, 2SLS ESTIMATES

|  | Log of Monthly Earnings (ZAR) | | |
|---|---|---|---|
|  | 2SLS Estimates | | |
|  | Year of birth interacted with non-white binary indicator | | |
|  | (1) | (2) | (3) |
| Schooling | 0.23 | 0.25 | 0.27 |
|  | (0.31) | (0.26) | (0.25) |
| Birth Year | 0.83 | 0.38 | 0.08 |
|  | (5.48) | (2.73) | (2.79) |
| Birth Year Squared | -0.00 | -0.00 | -0.00 |
|  | (0.03) | (0.02) | (0.02) |
| Experience[a] | 0.28*** | 0.28*** | 0.27*** |
|  | (0.04) | (0.02) | (0.02) |
| Experience Squared | -0.01*** | -0.01*** | -0.01*** |
|  | (0.00) | (0.00) | (0.00) |
| Male (=1 if male) | 0.27 | 0.29* | 0.29* |
|  | (0.28) | (0.16) | (0.16) |
| Black (=1 if black) | -0.55 | -0.39 | -0.36 |
|  | (1.01) | (0.33) | (0.34) |
| Coloured[d] (=1 if coloured) | -0.21 | -0.12 | -0.07 |
|  | (0.82) | (0.36) | (0.36) |
| Ability[b] |  | 0.07 |  |
|  |  | (0.24) |  |
| Literacy Component[c] |  |  | -0.03 |
|  |  |  | (0.15) |
| Numeracy Component[c] |  |  | 0.07 |
|  |  |  | (0.13) |
| Constant | -32.96 | -15.11 | -3.47 |
|  | (219.36) | (106.80) | (109.16) |
| F-statistic | 1580.14 | 1580.14 | 1580.14 |
| $R^2$ statistic | 0.40 | 0.40 | 0.40 |
| Observations | 940 | 940 | 940 |

*Notes:* Cape Area Panel Study. Earnings is in Constant 2002 ZARs. Exchange rate from 2002 is 10.539 USD-ZAR. (a) Experience is calculated by taking the difference of one's age and one's schooling minus six years following Mincer (1974), Boissiere, Knight and Sabot (1985) and Lemieux (2006). (b) Aggregate ability is calculated using principal component analysis on the Literacy and Numeracy Evaluation survey questions. (c) Components of Literacy and Numeracy Evaluation are calculated using principal component analysis on two sets of survey questions: literacy-related questions and numeracy-related questions. Restrictions are categorized by age or both age and grade. The age restriction is a minimum age of 15 to capture the school-leaving age. The grade restriction is a maximum schooling attainment of 9th grade, or the equivalent of school-leaving attainment. This eliminates learners who continued to study past the minimum school-leaving age, and are unlikely to be affected by the instrument as per Angrist and Krueger (1991). (d) Coloured is the term for a person of mixed European ("white") and African ("black") or Asian ancestry, as officially defined by the South African government from 1950 to 1991 and by the current national statistical bureau of South Africa (called Statistics SA) (Statistics SA 2017). Standard errors are clustered at the neighborhood level. Standard errors are clustered at the neighborhood level.
*** Significant at the 1 percent level.
** Significant at the 5 percent level.
* Significant at the 10 percent level.

# ONLINE APPENDIX B

For consistency, we follow the methodological approach used by Hansen et al. (2004), Cascio and Lewis (2006), and Segal (2012) to adjust evaluation measures according to age disparities. Specifically, we adjust the total evaluation score and the numeracy and literacy evaluation measures by regressing the original CAPS Wave 1 evaluation measures on dummy variables for schooling and age at the time of the Wave 1 evaluations. We then normalize the evaluation measures. Using this reconstructed measure of cognitive evaluations, we re-estimate specifications (1) through (4), which serve as the basis for the results reported in Table 5 through to Table 10. Tables B1 through B5 report the results based on the reconstructed cognitive evaluation measures.

TABLE B1. CAPS SURVEY (WAVE 5): RETURNS TO EDUCATION WITH RECONSTRUCTED ABILITY MEASURE(S), OLS ESTIMATES

|  | Log of Monthly Earnings (ZAR) | | |
|---|---|---|---|
|  |  | With Ability[b] | With Specific Ability Domains[c] |
|  | (1) | (2) | (3) |
| Schooling (in years) | 0.14*** | 0.13*** | 0.13*** |
|  | (0.01) | (0.01) | (0.01) |
| Experience[a] | 0.05*** | 0.06*** | 0.06*** |
|  | (0.02) | (0.02) | (0.02) |
| Experience Squared | 0 | -0.00* | -0.00** |
|  | (0.00) | (0.00) | (0.00) |
| Male (=1 if male) | 0.19*** | 0.18*** | 0.17*** |
|  | (0.03) | (0.03) | (0.03) |
| Black (=1 if black) | -0.60*** | -0.52*** | -0.50*** |
|  | (0.10) | (0.10) | (0.11) |
| Coloured[c] (=1 if coloured) | -0.15 | -0.1 | -0.07 |
|  | (0.10) | (0.10) | (0.10) |
| Aggregate Ability |  | 0.07*** |  |
|  |  | (0.02) |  |
| Literacy Component |  |  | 0.01 |
|  |  |  | (0.02) |
| Numeracy Component |  |  | 0.08*** |
|  |  |  | (0.02) |
| Constant | 5.95*** | 5.92*** | 5.89*** |
|  | (0.14) | (0.14) | (0.14) |
| $R^2$ statistic | 0.27 | 0.27 | 0.27 |
| Observations | 1,973 | 1,973 | 1,973 |

*Notes*: Cape Area Panel Study. Earnings is in Constant 2002 ZARs. Exchange rate from 2002 is 10.539 USD-ZAR. (a) Experience is calculated by taking the difference of one's age and one's schooling minus six years following Mincer (1974), Boissiere, Knight and Sabot (1985) and Lemieux (2006). (a) Aggregate ability is calculated using principal component analysis on the Literacy and Numeracy Evaluation survey questions. (b) Components of Literacy and Numeracy Evaluation are calculated using principal component analysis on two sets of survey questions: literacy-related questions and numeracy-related questions. (c) Coloured is the term for a person of mixed European ("white") and African ("black") or Asian ancestry, as officially defined by the South African government from 1950 to 1991 and by the current national statistical bureau of South Africa (called Statistics SA) (Statistics SA 2017). Standard errors are clustered at the neighborhood level.
*** Significant at the 1 percent level.
** Significant at the 5 percent level.
* Significant at the 10 percent level

TABLE B2. HAALSI SURVEY: RETURNS TO EDUCATION WITH RECONSTRUCTED ABILITY MEASURE(S), OLS ESTIMATES

|  | Log of Monthly Earnings (ZAR) | | |
|---|---|---|---|
|  |  | With Ability[b] | With Specific Ability Domains[c] |
|  | (1) | (2) | (3) |
| Schooling (in years) | 0.10*** | 0.10*** | 0.10*** |
|  | (0.01) | (0.01) | (0.01) |
| Experience[a] | 0.03*** | 0.03*** | 0.03*** |
|  | (0.01) | (0.01) | (0.01) |
| Experience Squared | -0.00** | -0.00** | -0.00** |
|  | (0.00) | (0.00) | (0.00) |
| Male (=1 if male) | 0.92*** | 0.89*** | 0.90*** |
|  | (0.05) | (0.05) | (0.05) |
| Aggregate Ability |  | 0.13*** |  |
|  |  | (0.03) |  |
| Memory Component |  |  | 0.08*** |
|  |  |  | (0.03) |
| Numeracy Component |  |  | -0.01 |
|  |  |  | (0.04) |
| Attention Component |  |  | 0.02 |
|  |  |  | (0.03) |
| Orientation Component |  |  | 0.07** |
|  |  |  | (0.04) |
| Constant | 5.16*** | 5.25*** | 5.22*** |
|  | (0.20) | (0.20) | (0.20) |
| $R^2$ statistic | 0.25 | 0.25 | 0.25 |
| Observations | 2,252 | 2,252 | 2,252 |

*Notes*: Health and Aging in Africa: A Longitudinal Study of an INDEPTH Community in South Africa (HAALSI). Earnings is from 1991 onward and is in Constant 2002 ZARs. Exchange rate from 2002 is 10.539 USD-ZAR. Log earnings is calculated using earnings data in Rand. (a) Experience is calculated by taking the difference of one's age and one's schooling minus six years following Mincer (1974), Boissiere, Knight and Sabot (1985) and Lemieux (2006). (b) Aggregate ability is calculated using principal component analysis on the Cognition section of HAALSI. (c) Components of HAALSI Cognition section are calculated using principal component analysis on four sets of survey questions: memory-related questions, numeracy-related questions, attention-related questions, and orientation-related questions.
*** Significant at the 1 percent level.
** Significant at the 5 percent level.
* Significant at the 10 percent level

TABLE B3. CAPS SURVEY (PANEL 1-3-4-5): RETURNS TO EDUCATION WITH RECONSTRUCTED ABILITY MEASURE(S), OLS ESTIMATES

|  | Log of Monthly Earnings (ZAR) | | |
|---|---|---|---|
|  |  | With Ability[b] | With Specific Ability Domains[c] |
|  | (1) | (2) | (3) |
| Schooling (in years) | 0.16*** | 0.15*** | 0.15*** |
|  | (0.02) | (0.02) | (0.02) |
| Experience[a] | 0.14*** | 0.15*** | 0.15*** |
|  | (0.03) | (0.03) | (0.03) |
| Experience Squared | -0.01*** | -0.01*** | -0.01*** |
|  | (0.00) | (0.00) | (0.00) |
| Male (=1 if male) | 0.24*** | 0.24*** | 0.23*** |
|  | (0.08) | (0.07) | (0.07) |
| Black (=1 if black) | -0.56*** | -0.44** | -0.41** |
|  | (0.17) | (0.18) | (0.18) |
| Coloured[d] (=1 if coloured) | -0.22 | -0.15 | -0.11 |
|  | (0.14) | (0.14) | (0.14) |
| Aggregate Ability |  | 0.10*** |  |
|  |  | (0.04) |  |
| Literacy Component |  |  | -0.03 |
|  |  |  | (0.05) |
| Numeracy Component |  |  | 0.13** |
|  |  |  | (0.05) |
| Wave 3 Dummy | 0.49*** | 0.50*** | 0.50*** |
|  | (0.07) | (0.07) | (0.07) |
| Wave 4 Dummy | 0.55*** | 0.56*** | 0.55*** |
|  | (0.07) | (0.07) | (0.07) |
| Wave 5 Dummy | 0.50*** | 0.52*** | 0.52*** |
|  | (0.10) | (0.10) | (0.10) |
| Constant | 4.97*** | 4.95*** | 4.91*** |
|  | (0.24) | (0.24) | (0.24) |
| $R^2$ statistic | 0.41 | 0.42 | 0.43 |
| Observations | 940 | 940 | 940 |

*Notes*: Cape Area Panel Study. Earnings is in Constant 2002 ZARs. Exchange rate from 2002 is 10.539 USD-ZAR. (a) Experience is calculated by taking the difference of one's age and one's schooling minus six years following Mincer (1974), Boissiere, Knight and Sabot (1985) and Lemieux (2006). (b) Aggregate ability is calculated using principal component analysis on the Literacy and Numeracy Evaluation survey questions. (c) Components of Literacy and Numeracy Evaluation are calculated using principal component analysis on two sets of survey questions: literacy-related questions and numeracy-related questions. Wave dummies work as follows: wave 3 dummy=1 if observation is from wave 3. wave 4 dummy=1 if observation is from wave 4. wave 5 dummy=1 if observation is from wave 5. (d) Coloured is the term for a person of mixed European ("white") and African ("black") or Asian ancestry, as officially defined by the South African government from 1950 to 1991 and by the current national statistical bureau of South Africa (called Statistics SA) (Statistics SA 2017). Standard errors are clustered at the neighborhood level.
\*\*\* Significant at the 1 percent level.
\*\* Significant at the 5 percent level.
\* Significant at the 10 percent level

TABLE B4. CAPS SURVEY: RETURNS TO EDUCATION, SECOND-STAGE ESTIMATION WITH RESTRICTIONS. BASED ON RECONSTRUCTED ABILITY MEASURE(S), 2SLS ESTIMATES

| Variables | Log of Monthly Earnings (ZAR) | | |
|---|---|---|---|
| | 2SLS Estimates | | |
| | (1) | (2) | (3) |
| Schooling | 0.28 | 0.29 | 0.29 |
| | (0.19) | (0.19) | (0.20) |
| Experience[a] | 0.27*** | 0.27*** | 0.28*** |
| | (0.03) | (0.03) | (0.03) |
| Experience Squared | -0.01*** | -0.01*** | -0.01*** |
| | (0.00) | (0.00) | (0.00) |
| Male (=1 if male) | 0.45*** | 0.45*** | 0.45*** |
| | (0.10) | (0.10) | (0.10) |
| Black (=1 if black) | 0.1 | 0.18 | 0.19 |
| | (0.59) | (0.60) | (0.60) |
| Coloured[d] (=1 if coloured) | 0.57 | 0.63 | 0.65 |
| | (0.58) | (0.58) | (0.59) |
| Ability[b] | | 0.06* | |
| | | (0.03) | |
| Literacy Component[c] | | | -0.01 |
| | | | (0.05) |
| Numeracy Component[c] | | | 0.08 |
| | | | (0.05) |
| Constant | 2.67 | 2.54 | 2.5 |
| | (1.96) | (1.97) | (2.00) |
| F-statistic | 3045.36 | 3045.36 | 3045.36 |
| $R^2$ statistic | 0.37 | 0.38 | 0.38 |
| Observations | 401 | 401 | 401 |

*Notes*: Cape Area Panel Study. Earnings is in Constant 2002 ZARs. Exchange rate from 2002 is 10.539 USD-ZAR. (a) Experience is calculated by taking the difference of one's age and one's schooling minus six years following Mincer (1974), Boissiere, Knight and Sabot (1985) and Lemieux (2006). (b)Aggregate ability is calculated using principal component analysis on the Literacy and Numeracy Evaluation survey questions. [2] Components of Literacy and Numeracy Evaluation are calculated using principal component analysis on two sets of survey questions: literacy-related questions and numeracy-related questions. (c) Restrictions are categorized by age or both age and grade. The age restriction is a minimum age of 15 to capture the school-leaving age. The grade restriction is a maximum schooling attainment of 9[th] grade, or the equivalent of school-leaving attainment. This eliminates learners who continued to study past the minimum school-leaving age, and are unlikely to be affected by the instrument as per Angrist and Krueger (1991). (d) Coloured is the term for a person of mixed European ("white") and African ("black") or Asian ancestry, as officially defined by the South African government from 1950 to 1991 and by the current national statistical bureau of South Africa (called Statistics SA) (Statistics SA 2017). Standard errors are clustered at the neighborhood level.
*** Significant at the 1 percent level.
** Significant at the 5 percent level.
* Significant at the 10 percent level.

TABLE B5. HAALSI AND CAPS WAVE 5 SURVEYS: RETURNS TO SCHOOLING BY DEGREE LEVEL. BASED ON RECONSTRUCTED ABILITY MEASURE(S), OLS ESTIMATES

| Variables | Log of Monthly Earnings (ZAR) | | | | | |
|---|---|---|---|---|---|---|
| | HAALSI | | | CAPS Wave 5 | | |
| | With Ability[b] | With Specific Ability Domains[c] | | With Ability[d] | With Specific Ability Domains[e] | |
| | (1) | (2) | (3) | (4) | (5) | (6) |
| Primary (7-9) | 0.35*** | 0.38*** | 0.38*** | 0.13* | 0.13* | 0.13* |
| | (0.07) | (0.07) | (0.07) | (0.07) | (0.07) | (0.07) |
| Lower Secondary (10-11) | 0.58*** | 0.60*** | 0.60*** | 0.33*** | 0.31*** | 0.31*** |
| | (0.09) | (0.09) | (0.09) | (0.08) | (0.07) | (0.08) |
| Upper Secondary (12 or more) | 1.40*** | 1.41*** | 1.42*** | 0.77*** | 0.74*** | 0.74*** |
| | (0.08) | (0.08) | (0.08) | (0.08) | (0.08) | (0.08) |
| Experience[a] | 0.04*** | 0.04*** | 0.04*** | 0.03 | 0.04** | 0.04** |
| | (0.01) | (0.01) | (0.01) | (0.02) | (0.02) | (0.02) |
| Experience Squared | -0.00*** | -0.00*** | -0.00*** | -0.00 | -0.00 | -0.00* |
| | (0.00) | (0.00) | (0.00) | (0.00) | (0.00) | (0.00) |
| Male (=1 if male) | 0.96*** | 0.90*** | 0.91*** | 0.17*** | 0.16*** | 0.15*** |
| | (0.05) | (0.05) | (0.05) | (0.03) | (0.03) | (0.03) |
| Black (=1 if black) | | | | -0.67*** | -0.57*** | -0.54*** |
| | | | | (0.11) | (0.11) | (0.11) |
| Coloured[e] (=1 if coloured) | | | | -0.24** | -0.18* | -0.15 |
| | | | | (0.11) | (0.11) | (0.11) |
| Aggregate Ability | | 0.20*** | | | 0.08*** | |
| | | (0.03) | | | (0.02) | |
| Literacy Component | | | | | | 0.01 |
| | | | | | | (0.02) |
| Numeracy Component | | | 0.10*** | | | 0.09*** |
| | | | (0.03) | | | (0.02) |
| Memory Component | | | 0.04 | | | |
| | | | (0.04) | | | |
| Attention Component | | | 0.01 | | | |
| | | | (0.03) | | | |
| Orientation Component | | | 0.11*** | | | |
| | | | (0.04) | | | |
| Constant | 5.31*** | 5.37*** | 5.35*** | 7.19*** | 7.11*** | 7.07*** |
| | (0.20) | (0.19) | (0.20) | (0.13) | (0.13) | (0.14) |
| $R^2$ statistic | 0.25 | 0.27 | 0.27 | 0.25 | 0.26 | 0.26 |
| Observations | 2,252 | 2,252 | 2,252 | 1,973 | 1,973 | 1,973 |

*Notes*: Cape Area Panel Study and HAALSI. Earnings is in Constant 2002 ZARs. Exchange rate from 2002 is 10.539 USD-ZAR. (a) Experience is calculated by taking the difference of one's age and one's schooling minus six years following Mincer (1974), Boissiere, Knight and Sabot (1985) and Lemieux (2006). (b) Aggregate ability is calculated using principal component analysis on the Cognition section of the HAALSI survey. (c) Components are calculated using principal component analysis on four sets of survey questions pertaining to each cognitive domain. (c) Aggregate ability is calculated using principal component analysis on the Literacy and Numeracy Evaluation survey questions. (d) Components of Literacy and Numeracy Evaluation are calculated using principal component analysis on two sets of survey questions: literacy-related questions and numeracy-related questions. (e) Coloured is the term for a person of mixed European ("white") and African ("black") or Asian ancestry, as officially defined by the South African government from 1950 to 1991 and by the current national statistical bureau of South Africa (called Statistics SA) (Statistics SA 2017). For the CAPS estimations reported in columns (4)-(6), standard errors are clustered at the neighborhood level.
\*\*\* Significant at the 1 percent level.
\*\* Significant at the 5 percent level.
\* Significant at the 10 percent level.